\documentclass{emulateapj}
\newcommand{\rxte}{{\it RXTE}}
\newcommand{\xte}{{\it RXTE}}

\newcommand{\tfe}{1E~1048.1--5937}
\newcommand{\tfn}{1E~2259$+$586}

\newcommand{\soe}{1RXS~J1708--4009}
\newcommand{\oft}{4U~0142$+$61}
\newcommand{\efo}{1E~1841--045}
\newcommand{\axj}{AX~J1845--0258}
\newcommand{\ett}{XTE~J1810--197}

\newcommand{\sfs}{CXO~J164710.2$-$455216}
\newcommand{\ffs}{1E~1547.0$-$5408}

\def\lapp{\ifmmode\stackrel{<}{_{\sim}}\else$\stackrel{<}{_{\sim}}$\fi}
\def\gapp{\ifmmode\stackrel{>}{_{\sim}}\else$\stackrel{>}{_{\sim}}$\fi}
\def\nuddd{\ifmmode\stackrel{\bf \,{\ldots} }{\textstyle
    \nu}\else$\stackrel{\,{\ldots} }{\textstyle \nu}$\fi}
\def\nudddd{\ifmmode\stackrel{\bf \,{\ldots} .}{\textstyle
    \nu}\else$\stackrel{\,{\ldots} .}{\textstyle \nu}$\fi}
\shorttitle{Activity from AXP 4U~0142+61}
\shortauthors{Gavriil et al.}
\begin{document}
\title{The 2006--2007 Active Phase of Anomalous X-ray Pulsar  4U~0142+61: Radiative and Timing Changes, Bursts, and Burst Spectral Features}

\author{Fotis~P.~Gavriil\altaffilmark{1} \altaffilmark{2},
        Rim~Dib\altaffilmark{3},
	Victoria~M.~Kaspi\altaffilmark{3}
	}
\altaffiltext{1}{NASA Goddard Space Flight Center, Astrophysics Science
                 Division, Code 662, Greenbelt, Maryland, 20771, USA}
\altaffiltext{2}{Center for Research and Exploration in Space Science and Technology, University of Maryland Baltimore County, 1000 Hilltop Circle, Baltimore,
                 Maryland, 21250, USA}
\altaffiltext{3}{Department of Physics, McGill University,
                 Montreal, QC, H3A~2T8, Canada.}

\begin{abstract}
After at least 6 years of quiescence, Anomalous X-ray Pulsar (AXP)
4U~0142+61 entered an active phase in 2006 March that lasted several
months and included six X-ray bursts as well as many changes in the
persistent X-ray emission.  The bursts, the first seen from this AXP
in $>$11 years of \textit{Rossi X-ray Timing Explorer} monitoring, all
occurred in the interval between 2006 April~6 and 2007 February~7.
The burst durations ranged from 0.4$-$1.8$\times$10$^{3}$~s.  The
first five burst spectra are well modeled by blackbodies, with
temperatures $kT\sim2-9$~keV. However, the sixth burst had a
complicated spectrum that is well characterized by a blackbody plus
two emission features whose amplitude varied throughout the burst. The
most prominent feature was at 14.0~keV.  Upon entry into the active
phase the pulsar showed a significant change in pulse morphology and a
likely timing glitch.  The glitch had a total frequency jump of
(1.9$\pm$0.4$)\times$10$^{-7}$~Hz, which recovered with a decay time
of 17$\pm$2~days by more than the initial jump, implying a net
spin-down of the pulsar.  Within the framework of the magnetar model,
the net spin-down of the star could be explained by regions of the
superfluid that rotate slower than the rest. The bursts, flux
enhancements, and pulse morphology changes can be explained as arising
from crustal deformations due to stresses imposed by the highly
twisted internal magnetic field. However, unlike other AXP outbursts,
we cannot account for a major twist being implanted in the
magnetosphere.

\end{abstract}


\section{Introduction}

It is now generally accepted that the class of objects referred to as
``Anomalous X-ray Pulsars'' (AXPs) are magnetars -- young isolated
neutron stars powered by the evolution of their high magnetic fields
\citep{td95, td96a}.  AXPs are X-ray pulsars with periods in the range
2--12~s, and X-ray luminosities ($\sim 10^{33}-10^{35}$~erg~s$^{-1}$)
that cannot be accounted for by their available spin-down energy.  The
magnetar model was first proposed to explain the dramatic behavior
exhibited by an apparently different object class -- the Soft Gamma
Repeaters (SGRs). SGRs show persistent properties similar to AXPs, but
they were first discovered by their enormous bursts of soft gamma rays
($> 10^{44}$~erg) and their much more frequent, shorter, and thus less
energetic bursts of hard X-rays. To date, SGR-like X-ray bursts have
been observed from six AXPs, thus solidifying the connection between
the two source classes \citep{gkw02,kgw+03,wkg+05,kbc+06, mgw+09,
  ks10}. For reviews of magnetar candidates and AXPs see \citet{wt06},
\citet{kas07}, and \citet{mer08}.

Thus far, only the magnetar model can explain the bursts observed from
SGRs and AXPs \citep{td95}. The internal magnetic field exerts
stresses on the crust which can lead to large scale rearrangements of
the external field, which we observe as giant flares. If the stress is
more localized, then it can fracture the crust and displace the
footpoints of the external magnetic field which results in short X-ray
bursts.  The highly twisted internal magnetic field also slowly twists
up the external field; magnetospheres of magnetars may therefore be
globally twisted \citep{tlk02}.  Reconnection in this twisted
magnetosphere has also been proposed as an additional mechanism for
the short bursts \citep{lyu02}.

In addition to bursts, AXPs and SGRs exhibit pulsed and persistent
flux variations on multiple different timescales.  An hours-long
increase in the pulsed flux has been seen to follow a burst in AXP
\tfe\ \citep{gkw06}.  On longer timescales, AXPs can exhibit abrupt
increases in flux which decay on several-week timescales. These occur
in conjunction with bursts and have been suggested as being due to
thermal radiation from the stellar surface after the deposition of
heat from bursts. Such flux enhancements have been observed in SGRs
\citep[see][for example]{wkg+01}. The flux enhancement of AXP \tfn\
during its 2002 outburst can also interpreted as burst afterglow
\citep{wkt+04}, however, a magnetospheric interpretation is also
plausible \citep{zkw+08}.  AXP \tfe\ exhibited three unusual flux
`flares.'  In the first two, the pulsed flux rose on week-long
timescales and subsequently decayed back on time scales of months
\citep{gk04,tgd+08}.  These variations have been tentatively
attributed to twists implanted in the external magnetosphere from
stresses on the crust imposed by the internal magnetic field.  AXPs
\ett, \ffs, and the AXP candidate \axj\ have also exhibited large flux
variations \citep{ims+04,hgr+08,gv98,tkgg06}, however it is not clear
whether these were of the abrupt rise type as in \tfn\ or the
slow-rise type as in \tfe.  \sfs\ showed a clear abrupt rise
\citep{mgc+07}.  \soe\ has been argued to have flux variations
associated with timing events \citep{igz+07}.  Finally, AXP \oft\ has
exhibited the longest timescale flux variations thus far, in which the
pulsed flux increased by 29$\pm$8\% over a period of 2.6 years
\citep{dkg07,gdk+10}.

4U~0142+61 is an 8.7-s AXP. It has a period derivative of
$\dot{P}$~=~0.2~$\times~10^{-11}$, implying a surface dipole magnetic
field of 1.3~$\times~10^{14}$~G.  4U~0142+61 was monitored by
{\emph{RXTE}} in 1997 and from 2000 to 2007.  \citet{gk02} showed that
\oft\ generally rotates with high stability.  \citet*{mks05} reported
a possible timing glitch in 1999 on the basis of an \textit{Advanced
Satellite for Cosmology and Astrophysics} (\textit{ASCA}) observation
in which the value of the frequency was marginally discrepant with the
frequency as reported by \citet{gk02}.  \citet{dkg07} showed that the
glitch may have occurred but is not required by the existing data.
\citet{dkg07} also reported on the evolution of the properties of
4U~0142+61 from 2000 March to 2006 April. In particular they reported
stable timing, and an evolution of the pulse profile in 2$-$4~keV
where the dip between the two peak was rising between 2000 and
2006. They also reported a 29$\pm$8\% increase in the pulsed flux
between 2002 May and 2004 December.  As of 2006 March, in the
published flux history of this source, there had been no reports of
any X-ray activity such as bursts or flares, as described above for
other AXPs.

Here we report on the first detection of bursts from AXP \oft, making
this the seventh AXP for which this phenomenon has been observed. We
also report that the source appears to have entered an active phase in
2006 March in which almost every aspect of the emission changed. Our
observations are described in Section~\ref{sec:observations}. Our
burst, pulsed morphology, pulse phase, pulsed flux, and timing
analysis are presented, respectively, in Sections~\ref{sec:bursts},
\ref{sec:profiles}, \ref{sec:phases}, \ref{sec:flux},
and~\ref{sec:timing}. In Section~\ref{sec:discussion}, we discuss the
possible origins of this behavior and the implications for the
magnetar model.

\section{Observations}
\label{sec:observations}

\begin{figure}
\centerline{\includegraphics[width=\columnwidth]{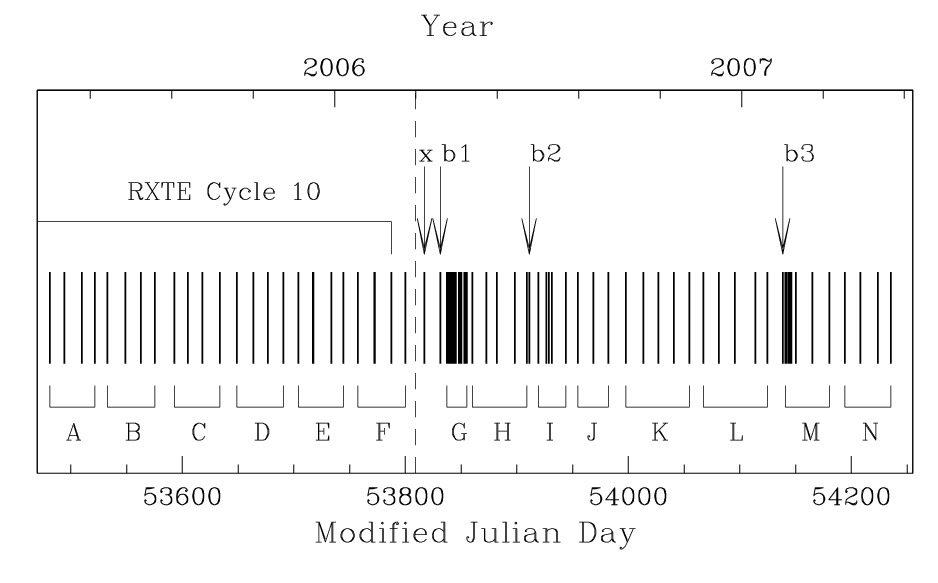}}
\caption
{ Epochs of the {\em{RXTE}} observations of 4U~0142+61 analyzed in
this paper (see \S~\ref{sec:profiles}).  Each observation is
represented by a vertical line.  The dashed line marks the entry of
the source into the active phase. An ``x'' marks the first observation
where the pulse profile was significantly different from the long-term
average. ``b1'', ``b2'', and ``b3'' mark the observations containing
bursts. To the left of the dotted line, the letters refer to groups of
observations having similar total integration times. To the right, the
letters refer to groups of observations having similar pulse profiles.
\label{fig:observations}
}
\end{figure}

4U~0142+61 has been monitored with 
\textit{Rossi X-ray Timing Explorer} (\xte) in 1997 and from 2000 to
2007. It has been monitored bi-monthly since 2005 March, with a
typical observation length of $\sim$5~ks. On 2006 March 23, the source
entered an active phase where many pulsed flux, spectral, and pulse
profile changes were observed. We detected 6 bursts in three
observations after the entry into the active phase. After each burst
detection, several {\emph{RXTE}} target-of-opportunity (ToO)
observations were made in addition to the regular monitoring.

Here, we present a detailed analysis of the three observations
containing bursts (see Table~\ref{tab:bursts}). We also present an analysis of
91 other observations spanning the 2005 March 21 to 2007 May 15 time period
(MJD 53481 to MJD 54235, Observation IDs 91070-05-04-00 to 92006-05-32-00).
32 of these observations were prior to the entry into the active phase, and
the remaining 59 observations were after. This long-term analysis was
performed in order to track the evolving pulsed flux and timing properties
of the source. 

\begin{deluxetable*}{l|c|cccc|c}
\tablewidth{2.0\columnwidth}
\tablecaption{4U 0142+61 Burst Observation Log and Burst Properties \label{tab:bursts}}
\startdata
\hline\hline
\multicolumn{7}{c}{Observations Containing Bursts}\\
\hline
Observation ID                           & 92006-05-03-00           & \multicolumn{4}{c|}{92006-05-09-01}                                                                       & 92006-05-25-00\\
Burst Date                               & 6 April 2006             & \multicolumn{4}{c|}{26 June 2006}                                                                         & 7 February 2007\\
Burst MJD (UTC)                          & 53831                    & \multicolumn{4}{c|}{53911}                                                                                & 54138\\
Number of Active PCUs                    & 3\tablenotemark{a}       & \multicolumn{4}{c|}{3}                                                                                    & 2\\
Number of Bursts                         & 1                        & \multicolumn{4}{c|}{4}                                                                                    & 1\\
\hline
\multicolumn{7}{c}{Burst Temporal and Spectral Properties\tablenotemark{b}}\\
\hline
Burst Number                             & 1                        & 2                         & 3                        & 4                        & 5                       & 6\\
$t_p$ \tablenotemark{c} (UT)                           & 07:09.55.544(7)          & 01:15:54.555(7)          & 01:15:55.119(58)         & 01:16:9.216(7)           & 01:20:0.131(3)          & 10:04:43.264(60)\\
$\tau_r$ \tablenotemark{d,i}(ms)                           & 7$^{+3}_{-2}$            & 7$^{+2}_{-2}$            & 58$^{+19}_{-16}$         & 7$^{+3}_{-4}$            & 3$^{+1}_{-2}$           & 60$^{+9}_{-2}$\\
$\tau_f$ \tablenotemark{e,i} (ms)                           & 33$^{+13}_{-20}$            & 18$^{+10}_{-13}$            & 65$^{+32}_{-40}$         & 188$^{+118}_{-24}$            & 228$^{+207}_{-26}$           & 3621$^{+264}_{-75}$\\
$\tau_t$ \tablenotemark{f,i} (s)                           & 4.2(2)           & 0.22(4)            & 12.21(6)       & 191.6(4)            & 40.7(2)         & 824.9(8)\\
$f_p$ \tablenotemark{g,i} (counts/s/PCU)     & 543$^{+174}_{-258}$                  & 852$^{+273}_{-358}$                   & 151$^{+28}_{-39}$                  & 329$^{+194}_{-40}$                  & 246$^{+214}_{-27}$                 & 551$^{+39}_{-11}$ \\
$f_p$ \tablenotemark{g,i,k} ($10^{-10}$ erg/s/cm$^{2}$)    & 187$^{+60}_{-89}$                  & 144$^{+46}_{-61}$                   & 61$^{+11}_{-16}$                  & 31$^{+18}_{-4}$                  & 52$^{+45}_{-6}$                 & 66$^{+5}_{-1}$ \\

$f_t$ \tablenotemark{h,i} (counts/s/PCU)     & 13.3(8)              & 125(21)             & 25.5(1)         & 8.47(1)              & 7.14(3)               & 12.70(1)\\
$f_t$ \tablenotemark{h,i,k} ($10^{-10}$ erg/s/cm$^{2}$)     & 4.6(3)                 & 21(4)    & 10.38(4)   &  0.7992(9)  & 1.516(6) &1.518(1) \\
$F_{\mathrm{tot}}$ \tablenotemark{j} (counts/PCU)    & 78(4)                    & 47(7)                   & 328(2)                   & 1686(3)                   & 346(2)                  & 12460(13)\\
$F_{\mathrm{tot}}$ \tablenotemark{j,k} ($10^{-10}$ erg/cm$^{2}$)  & 27(1)  & 8(1) & 134(1)  & 159.1(2)  & 73.4(5) & 1490(2)\\
$T_{90}$ \tablenotemark{l}    (s)         & 8.4(4)      & 0.4(1)  & 27.5(1)    & 434(1)  & 86.7(6)  & 1757(2)\\      
Phase \tablenotemark{m}  (cycles)         & 0.503(7)                 & 0.487(4)                  & 0.54(2)                  & 0.173(4)                 & 0.749(4)                & 0.358(4) \\
$kT$ \tablenotemark{n} (keV)              & 8.8$^{+12.3}_{-3.8}$      & 5.4$^{+1.7}_{-1.2}$    & 4.8$^{+0.3}_{-0.3}$   & 2.4$^{+0.2}_{-0.2}$   & 2.7$^{+0.2}_{-0.2}$  & 2.5$^{+0.2}_{-0.2}$ \\
$R$ \tablenotemark{o} (m)                & 24$^{+20}_{-19}$                      & 163$^{+61}_{-50}$                      & 100$^{+8}_{-8}$                      & 107$^{+17}_{-15}$                      & 132$^{+16}_{-15}$                     &  151$^{+22}_{-20}$ \\
$\chi^2/{\mathrm{DoF}}$ \tablenotemark{p} [DoF]            & 0.78 [7]                      & 0.87 [3]                       & 1.09[38]                      & 0.49 [37]                     & 0.81 [32]                   & 1.00 [48] 
\enddata
\tablenotetext{a}{One PCU switched on and another one off part way through the observation, but the total number of active PCUs stayed constant throughout the observation.}
\tablenotetext{b}{All quoted errors represent 1-$\sigma$ uncertainties. In the fits, the column density was held fixed at $N_H$$=$0.64$\times$10$^{22}$~cm$^{-2}$ \citep{dv06b}.}
\tablenotetext{c}{Time of burst peak.}
\tablenotetext{d}{Rise Time.}
\tablenotetext{e}{Short-term decay time of the tail of the burst.}
\tablenotetext{f}{Long-term decay time of the tail of the burst.}
\tablenotetext{g}{2--60 kev peak flux.}
\tablenotetext{h}{2--60 keV flux in the tail component of the burst.}
\tablenotetext{i}{These parameters were obtained by   fitting the model given by Eq.~\ref{eq:model} to the burst.}  
\tablenotetext{j}{2--60~keV fluence. These values were obtained by using Eq.~\ref{eq:fluence}.}
\tablenotetext{k}{All quoted fluxes and fluences in CGS units are unabsorbed.} 
\tablenotetext{l}{$T_{90}$ duration, defined as the time it took to collect 90\% of the total burst fluence. These values were obtained by using Eq.~\ref{eq:T90}.}
\tablenotetext{m}{The phases are relative to the template shown in Figure~\ref{fig:phases}.}
\tablenotetext{n}{Blackbody temperature. For burst 6, the temperature is that of the blackbody component after accounting for the apparent spectral features (see \S\ref{sec:burstspectral}).}
\tablenotetext{o}{The blackbody radius was calculated assuming a distance to the source of 3.5~kpc \citep{dv06a}.}
\tablenotetext{p}{Reduced $\chi^2$ for the spectral fits.}
\end{deluxetable*}

Figure~\ref{fig:observations} shows a timeline of the 94 analyzed
observations.  Note that two segments of any observation that was
split counted as separate observations if the segments were given
different observation IDs.  The ranges of epochs with an increased
density of observations contain the ToO observations. Prior to the
active phase, the groups of observations referred to by capital
letters in the Figure have similar total integration times. The groups
of observations in the active phase have similar pulse profiles. These
groups will be referred to in Sections~\ref{sec:profiles}
and~\ref{sec:timing}.

\section{Analysis and Results} 

All data presented here are from the Proportional Counter Array
\citep[PCA;][]{jmr+06} aboard \xte.  The PCA is made up of five
identical and independent proportional counter units (PCUs). Each PCU
is a Xenon/methane proportional counter with a propane veto layer. The
data were collected in either \texttt{GoodXenonwithPropane} or
\texttt{GoodXenon} mode which record photon arrival times with
$\sim$1-$\mu$s resolution and bins them with 256 spectral channels in
the $\sim$2--60~keV band.
For all \xte\ observations of 4U~0142+61, photon arrival times at
each epoch were adjusted to the solar system barycenter using the
position obtained by \citet{pkw+03} from \textit{Chandra X-ray Observatory} data.

\subsection{Burst Analysis}
\label{sec:bursts}

For each monitoring observation of 4U~0142+61, using software that can
handle the raw telemetry data, we generated 31.25-ms lightcurves using
all Xenon layers and events in the 2--20~keV band. These lightcurves
were searched for bursts using our automated burst search algorithm
introduced in \citet{gkw02} and discussed further in \citet{gkw04}. In
an observation on 2006 April 6, we detected a significant burst, and
four more bursts were detected in a single observation on 2006 June
25. The sixth and most energetic burst was detected on 2007 February
7. The bursts were significant in each active PCU. See
Table~\ref{tab:bursts} for the number of active PCUs in each burst
observation, as well as for the burst observation epochs.

To analyze these bursts we created event lists in
FITS\footnote{\url{http://fits.gsfc.nasa.gov}} format using the
standard
\texttt{FTOOLS}\footnote{\url{http://heasarc.gsfc.nasa.gov/docs/software/ftools/}}. For
consistency with previous analyses of SGR/AXP bursts we extracted
events in the 2-60~keV band. The burst
lightcurves are displayed in Figure~\ref{fig:lc} (LEFT).  Our
techniques for characterizing the temporal and spectral properties of
bursts were discussed in detail in \citet{gkw04} but we repeat them
here as some required modification in order address the distinct
properties of these bursts.

Before measuring any burst parameters we determined the instrumental
background using the \texttt{FTOOL} \texttt{pcabackest}. We extracted
a background model lightcurve using the appropriate energy band and
number of PCUs. \texttt{pcabackest} determines the background count rate only on
16-s time intervals, so we interpolated these values by fitting a
polynomial of order 6 to the entire observation, which yielded a good
fit for each data set.

\subsubsection{Burst Temporal and Energetic Properties}
\label{sec:burststemporal}

\begin{figure*}
\plottwo{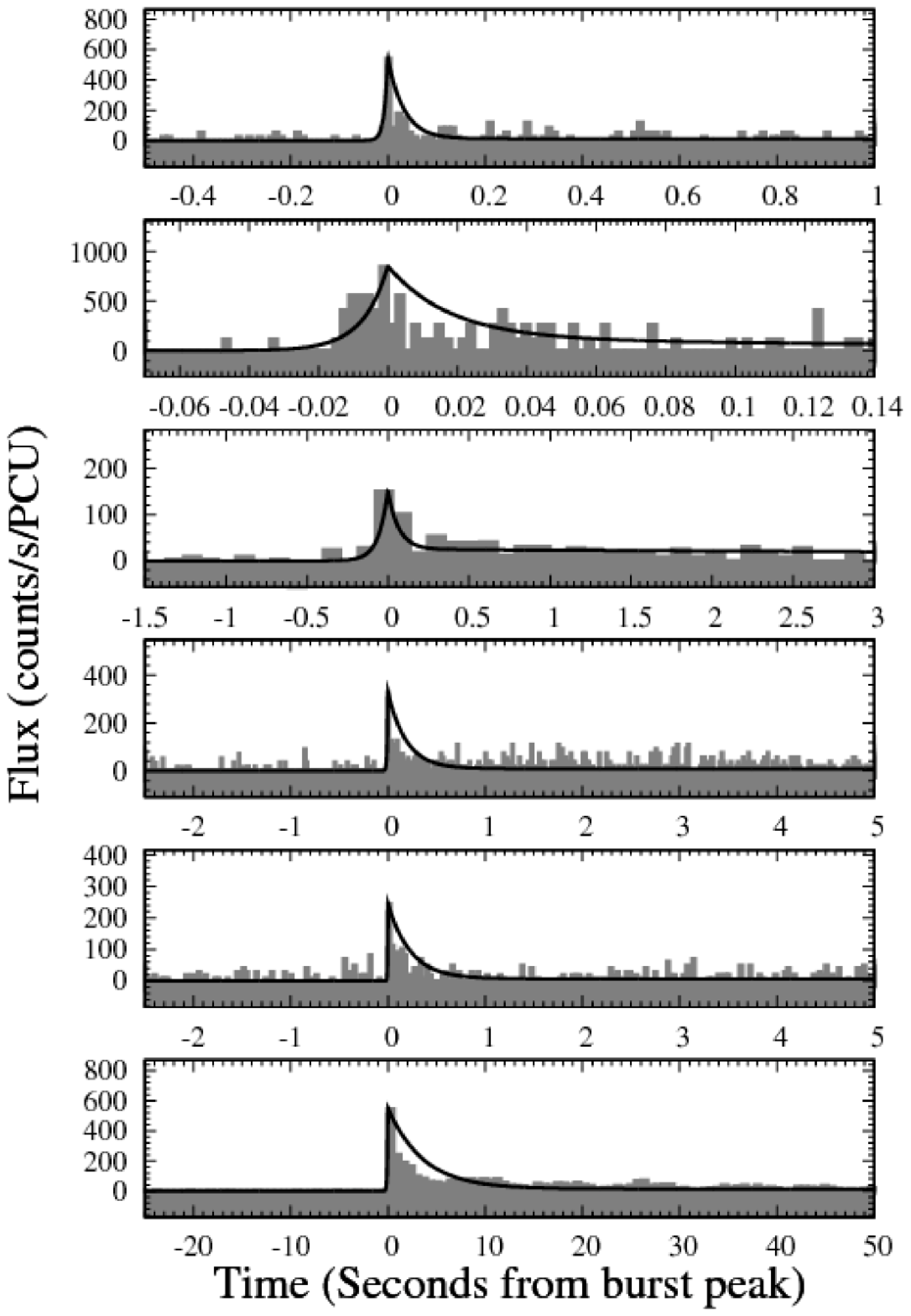}{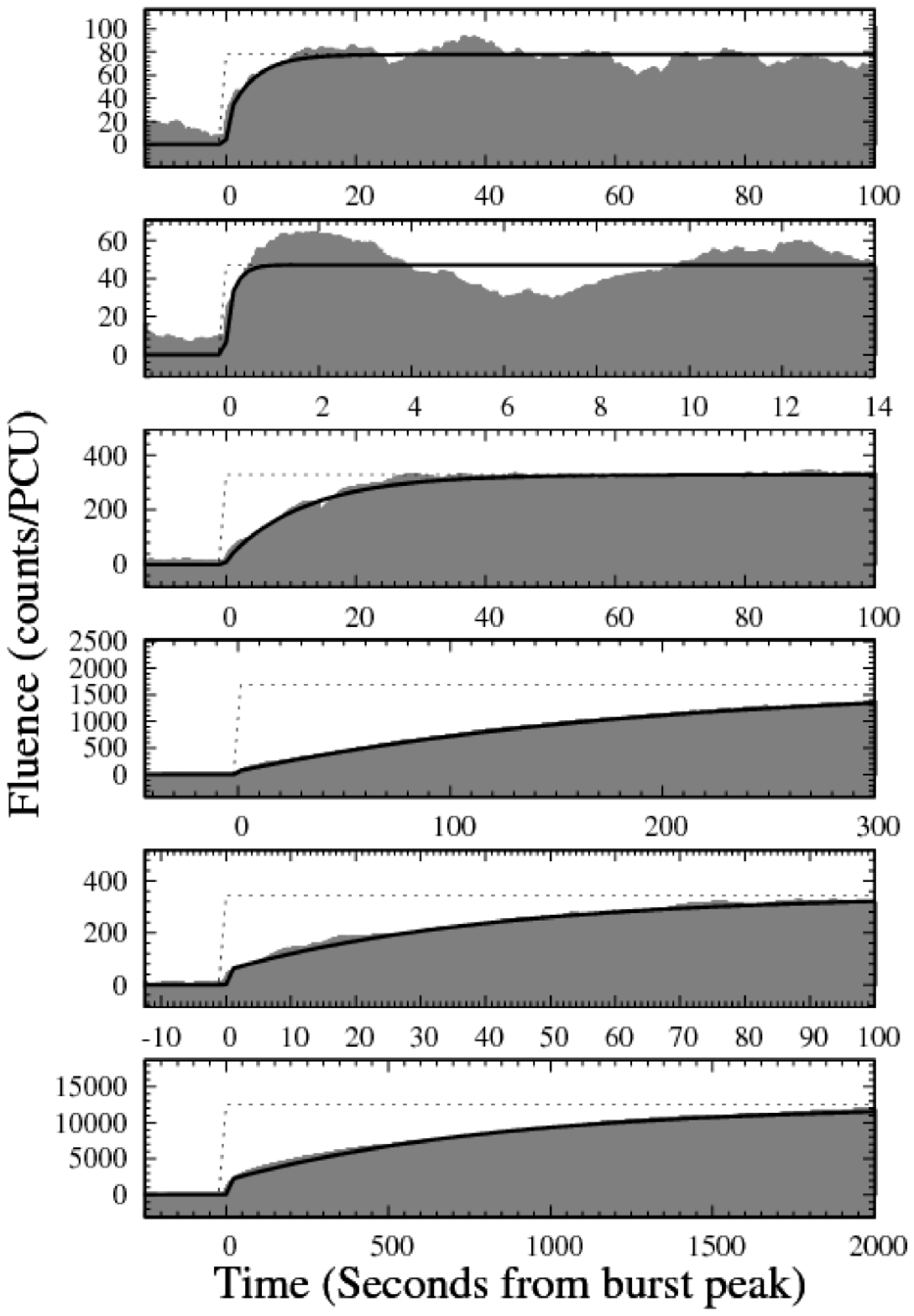}
\caption { LEFT -- The gray filled curves represent the 2--60~keV
  burst lightcurves as observed by \xte. The binning was chosen such
  that the flux in the peak is equal to the model peak flux, $f_p$ (see Eq.~\ref{eq:model}), as determined by maximum likelihood likelihood  testing. This binning scheme allow us resolve the bursts
  at the finest permissible time resolution. In descending order, the
  time bin widths are: 11.0, 2.3, 107.6, 18.3, 32.6, and 686.3 ms.The solid
  black curves are the best-fit exponential rise and decay models
  (Eq.~\ref{eq:model}).  RIGHT -- The gray filled curves represent the
  2-60 keV integrated burst lightcurves as observed by \xte. For each
  burst lightcurve we subtracted the instrumental background, the
  background due to other sources in the FOV, and the contribution of
  the tails of neighboring bursts before integrating.  Notice that
  some bursts have very long tails.  The solid black curve is the
  integral of the model in the left panel. The dashed curve is a step
  function whose height corresponds to the total flux as determined by
  Eq.~\ref{eq:fluence}.
\label{fig:lc}
}
\end{figure*}

The burst peak time ($t_p$) was  determined using the method described in \citet{gkw04, gkw06}.
Usually, to measure the fluence for SGR and AXP bursts, we subtract
the instrumental background for the lightcurve, integrate the light
curve and fit the result to a step function with a linear term whose slope is
the ``local'' background rate. The burst fluence in this case is the height
of the step function. Although this technique worked well for the
first burst, which was a short, isolated event, it was not appropriate
for bursts 2, 3, and 4 because they had overlapping tails, and bursts
4, 5 and 6 had tails that extended beyond the end of the
observation. Thus, we opted to determine the best-fit burst  profile , $f(t)$, consisting of 
exponential rises and decays,
\begin{equation}
f(t) = \left\{ \begin{array}{lr} 
f_p e^{\left(\frac{t-t_p}{\tau_r}\right)}     +  b(t)  + B   & t \le t_p \\ 
& \\
(f_p-f_t) e^{-\left(\frac{t-t_p}{\tau_f}\right)} + f_t e^{-\left(\frac{t-t_p}{\tau_t}\right)}    & t > t_p,\\
+b(t)  + B       & \\
\end{array} \right.
\label{eq:model}
\end{equation}
via maximum likelihood testing.  The burst peak time ($t_p$), and
the background model ($b(t)$) determined using \texttt{pcabackest} and
the ``local'' background ($B$) were held fixed at the values
determined by the methods outlined above.  We fit for the rise time
($\tau_r$), the peak flux ($f_p$), the flux of the long-term tail
component ($f_t$), and for the short ($\tau_f$) and long-term
($\tau_t$) decay times of the tail of the burst.   Figure~\ref{fig:lc} (LEFT)
shows our observed and model burst lightcurves. The binning was chosen such
that the flux in the peak bin was equal to the flux determined from
our maximum likelihood fits. We then integrated
our events to obtain fluence lightcurves. Figure~\ref{fig:lc}  
(RIGHT) shows our observed and model burst fluence time series.  For
both figures, the contribution from neighboring bursts has been
subtracted.  


As is done for $\gamma$-ray bursts and SGR and AXP bursts, we
characterized the burst duration by $T_{90}$, the time when 90\% of
the total burst fluence has been collected. From Eq.~\ref{eq:model} it
follows that the total fluence, $F_{\mathrm{tot}}$,  is given by
\begin{equation}
F_{\mathrm{tot}} = f_p \tau_r + (f_p - f_t)\tau_f + f_T\tau_t,
\label{eq:fluence}
\end{equation}
and the $T_{90}$ duration is given by 
\begin{equation}
T_{90} = \tau_t \log \left( {10f_t \tau_t}/{F_{\mathrm{tot}}} \right).
\label{eq:T90}
\end{equation}
Although Eq.~\ref{eq:fluence} is exact, Eq.~\ref{eq:T90} is an
approximation that holds very well  because $T_{90} \gg \tau_r, \tau_f$  for all
bursts. This fact is made obvious by Figure~\ref{eq:fluence} (RIGHT).  All
burst temporal parameters are presented in Table~\ref{tab:bursts}.

\subsubsection{Burst Spectral Properties}
\label{sec:burstspectral}

Burst spectra were extracted using all the counts within their
$T_{90}$ interval.  We then filtered the spectra to include counts
only from the Xenon layers and from all active PCUs other than PCU
0. PCU 0 was ignored because of the loss of its propane layer and due
to its frequent breakdowns.  Background intervals were extracted from
long, hand-selected, featureless intervals prior to the
bursts. Response matrices were created using the \texttt{FTOOL}
\texttt{pcarsp}. The burst spectra were grouped such that there were
at least 10 counts per bin after background subtraction. Burst
spectra, background spectra, and response matrices were then read into
the spectral fitting package
\texttt{XSPEC}\footnote{\url{http://xspec.gsfc.nasa.gov}} v12.3.1. The
spectra were fit to photoelectrically absorbed blackbodies using the
column density found by \citet{dv06b}. Only bins in the 2--30~keV
band were included in the fits. The blackbody model provided an
adequate fit for bursts 1 through 5; model parameters are presented in
Table~\ref{tab:bursts}.

\begin{figure}
\centerline{\includegraphics[width=\columnwidth]{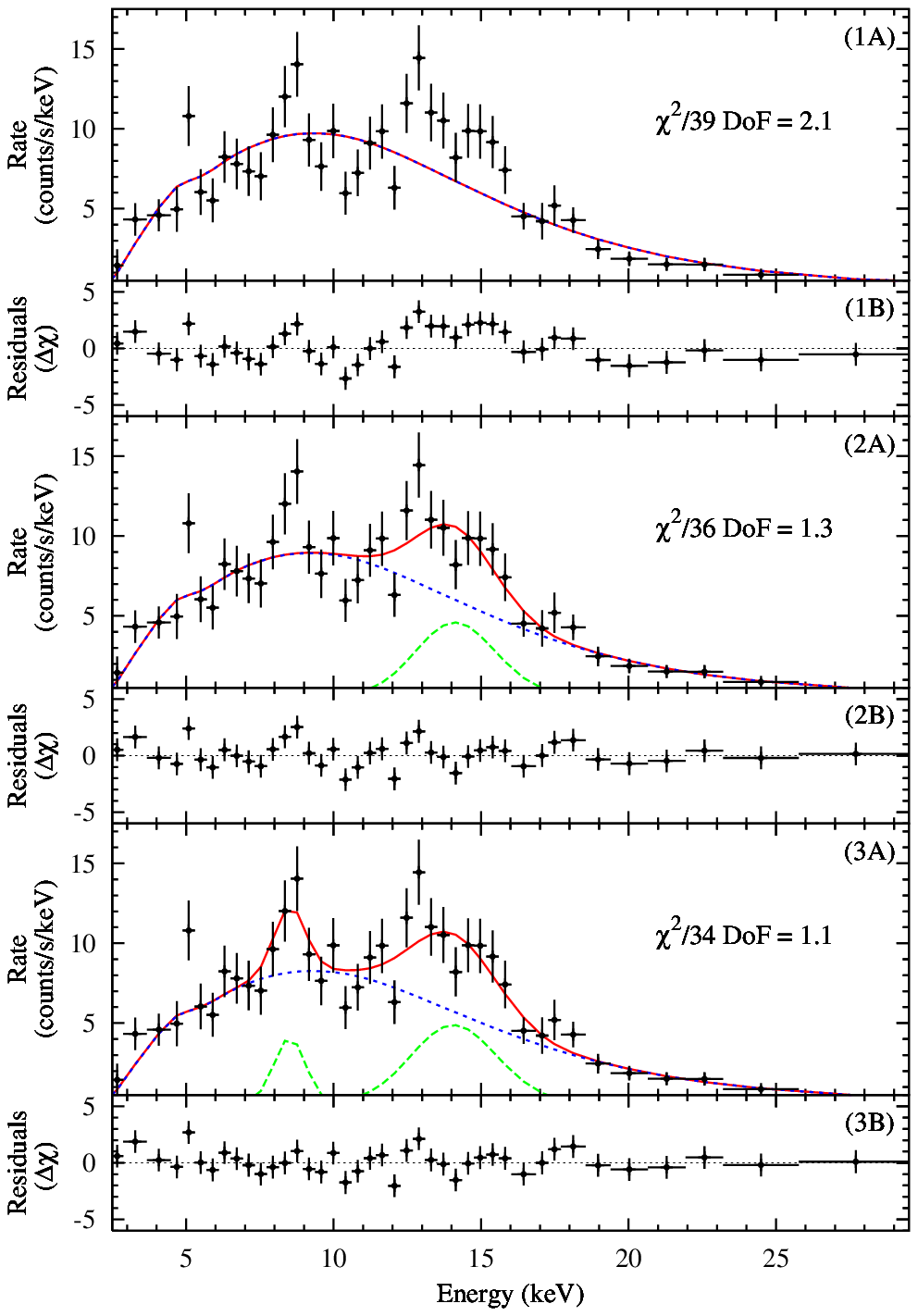}}
\caption {Panel 1A: The spectrum of burst 6 extracted over a pulse
  period (8.69~s) starting 50~ms before the burst peak.  The red solid
  lines represents the best-fit model-- a photoelectrically absorbed
  blackbody (red curve).  The fit had a reduced $\chi^2$ of 2.1 for 39
  degrees of freedom (DoF).  Panel 1B: The difference between the data
  and the model shown in Panel 1A, divided by the uncertainty.  Panels
  2A and 2B : Same, except here the best-fit model (red curve)
  consists of a blackbody (dashed blue curve) plus a Gaussian emission
  lines (dashed green curve), all photoelectrically absorbed. The fit
  had a reduced $\chi^2$ of 1.3 for 36 DoF.  Panels 3A and 3B: Same,
  except here the best-fit model (red curve) consists of a blackbody
  (blue curve) plus a two Gaussian emission lines (dashed green
  curves), all photoelectrically absorbed. The fit had a reduced
  $\chi^2$ of 1.1 for 34 DoF.
\label{fig:feature}}
\end{figure}

Extracted over its very long $T_{90}$ interval (see
Table~\ref{tab:bursts}), the spectrum of Burst 6 was well fit by a
simple blackbody, however, there were hints of spectral features. On
much shorter timescales, the spectrum of burst 6 was not well modeled
by any simple continuum model and the presence of spectral features
is clear (see Fig.~\ref{fig:feature}).  
We looked closely at the first 8.69~s (one pulse period) of the burst
(see Fig.~\ref{fig:feature}), and determined that a simple continuum
model with less than two spectral features was not a statistically
good representation of the data. A simple photoelectrically absorbed
blackbody yielded a reduced $\chi^2$ of 2.1 for 39 degrees of freedom
(DoF). Adding a single Gaussian emission lines improved the fit but
not sufficiently, yielding reduced $\chi^2$s of 1.3 for 36 DoF.  Only
a photoelectrically absorbed blackbody plus two Gaussian emission
lines provided an adequate fit, with reduced $\chi^2$$=$1.1 for 34
DoF. With this best-fit model, we find lines at energies
8.6$^{+0.1}_{-0.2}$, 14.2$^{+0.3}_{-0.3}$. Note, that 8.6~keV line was
very narrow, we thus kept its width fixed at the detectors energy
resolution, this was not necessary for the 14.2~keV line as it was
relatively broad.  Our results from fitting different models to the
first 8.69~s of the burst spectrum are summarized in
Figure~\ref{fig:feature}, and the parameters returned from our
best-fit model are listed in Table~\ref{tab:feature}.

We investigated whether different continuum models could describe the
feature rich burst spectrum just as well as a blackbody.  We tried a
simple power-law model, as well as more complicated continuum models
such as a Cutoff Power-law \citep[PLCUT;][]{wsh83} and a
Negative/Positive Exponential \citep[NPEX;][]{mih95}. The PLCUT and
NPEX models are often used to fit the continuum component of the line
rich spectra of accreting X-ray pulsars \citep[see][]{chr+02}. All
three of our alternate continuum models also required two emission
features at similar energies as the blackbody, however they did not
provide improved fits.  The best alternative to a blackbody was the
PLCUT model with two Gaussian emission lines at 8.8 and 14.3~keV with
reduced $\chi^2=$1.1 for 33 DoF. Notice how this model combination had
the same reduced $\chi^2$ as the blackbody combination, but required
one more free parameter.

We then determined whether the features could be described by
absorption rather than emission. We tried three different absorption
models: cyclotron absorption, Gaussian absorption (multiplicative
model), and simply subtracting Gaussians. We tested all three of these
absorption models with all the continuum models described above. No
permutation provided a better fit than a blackbody with two emission
lines, a simple power-law with two Gaussian absorption lines at 10.8~keV
and a broad (Gaussian width $\sigma=$4.4~keV) feature at 2.5~keV being
the best absorption model combination with reduced $\chi^2=1.2$ for 33
DoF.

\begin{figure*}
\centerline{\includegraphics[height=1.75\columnwidth, angle=270]{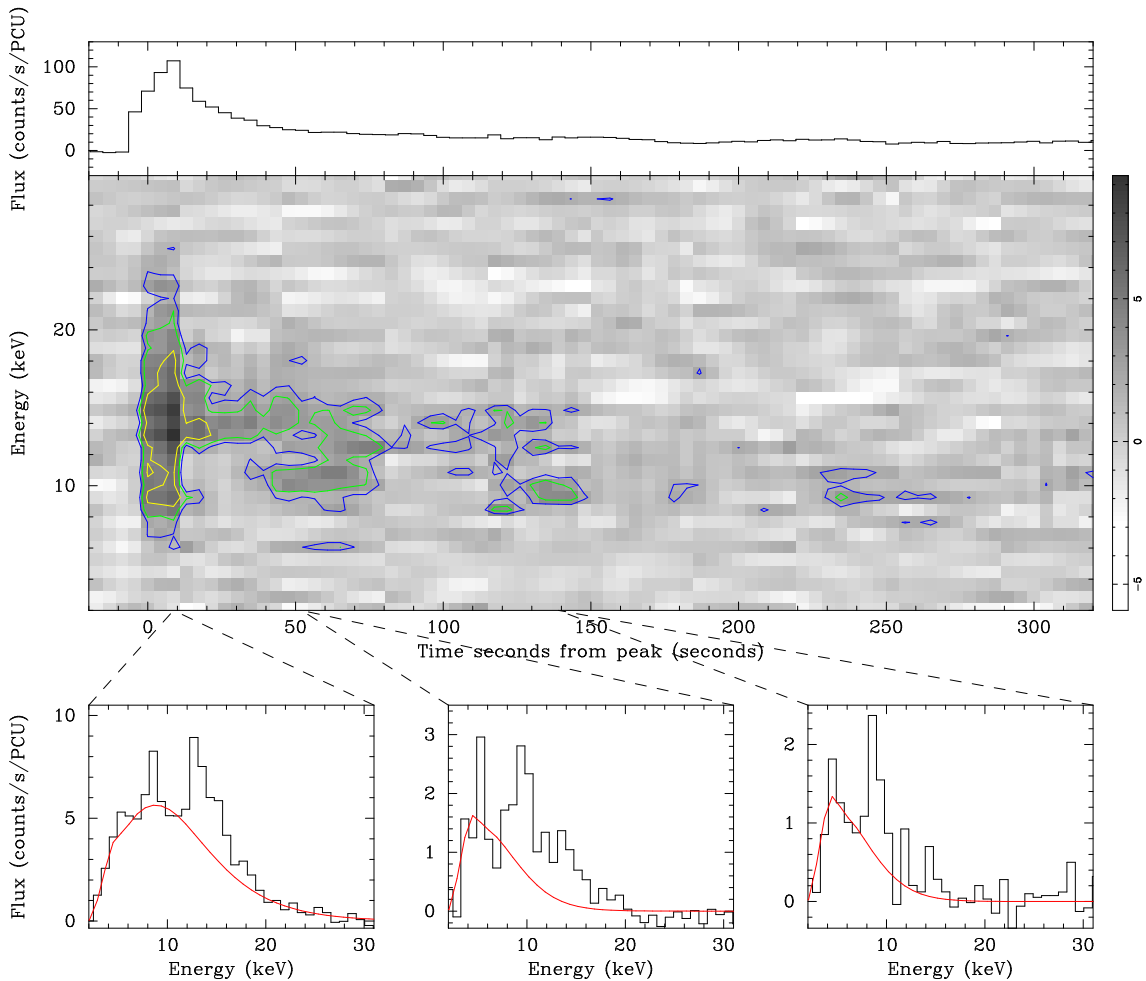}}
\caption {Spectral evolution of burst 6.  Top: Burst lightcurve binned
  with 4.34-s (half a pulse period)  time resolution and extracted over the same energy band
  used for spectral fitting, 2.5--30~keV. Middle: Surface plot of the
  difference between the burst spectrum and a simple continuum model
  (blackbody) in terms of change in $\chi^2$.  Each vertical slice in
  the surface plot corresponds to a spectrum that was calculated over
  a two-pulse-period long window. This window was translated across
  the burst in half-a-pulse-period long steps. For each spectrum, a
  best-fit blackbody model was subtracted.  The color wedge on the
  right maps the colors to the resulting change in $\chi$.  The
  contours indicate the regions where the spectrum differed by the
  model by 1-$\sigma$ (blue contour), 2-$\sigma$
  (green contour) and 3-$\sigma$ (yellow
  contour), respectively.  The feature at 14.2~keV is highly
  significant at the start of the burst, and remains detectable at the
  $>$1-$\sigma$ level for $\sim$130~s.  Notice, that the significance
  of the 8.6~keV feature is intermittent, becoming comparable to that
  of the 14.2~keV feature at later times, and is the only feature
  detectable at the $>$1-$\sigma$ beyond $\sim$130~s from the burst
  peak.  Bottom: The insets are examples of individual
  two-pulse-period long spectra used to generate the surface
  plot. These spectra were extracted at 8.69, 52.1 and 139.0~s from
  the burst onset, respectively. The red curves indicate the
  corresponding best-fit blackbody model.
  \label{fig:dynamic} }
\end{figure*}

To study the temporal
evolution of these features we extracted 17.38-s  (two pulse periods) long spectra in steps of
4.34~s (half a pulse period)  from the peak of the burst, and repeated the spectral fitting
procedure outlined above.  Fig.~\ref{fig:dynamic} displays the
spectral evolution of the features.  The middle panel is a surface
plot of the change in $\chi$ after subtracting the continuum model
(blackbody).  A vertical slice in the middle panel corresponds to a
spectrum of the burst extracted over a two-pulse-period long interval, and these spectra
were extracted in time steps of half a spin-period.  The features showed clear temporal
variability and were most prominent near the onset of the burst (see
Fig.~\ref{fig:dynamic}).
From the surface plot in Fig.~\ref{fig:dynamic} (middle panel),
we see that the 14.2~keV is highly significant at the start of the
burst, and remains detectable at the $>1$-$\sigma$ level for
$\sim$130~s. The significance of the 8.6~keV feature is intermittent, but
it remains detectable at the $>1$-$\sigma$ level beyond 130~s from the
burst onset.

\begin{deluxetable}{lc}
\tablewidth{\columnwidth}
\tablecaption
{
Burst 6 Spectral Fit
\label{tab:feature}
}
\tablehead{\colhead{Parameter} & \colhead{Value\tablenotemark{a}}}
\startdata
Column Density\tablenotemark{b}, $N_H$ ($10^{22}$~cm$^{-2}$) & 0.64 \\
2--60 keV  Flux\tablenotemark{c} (10$^{-10}$~erg~s$^{-1}$cm$^{-2}$) & 39.6$^{+3.7}_{-3.6}$\\   
Reduced $\chi^2 (\textrm{Degrees of Freedom})$ & 1.1(34) \\ 
\cutinhead{Blackbody Component}\\
Temperature, $kT$  (keV) & 5.8$^{+0.3}_{-0.3}$ \\
Radius\tablenotemark{c} (km) & 0.19$^{+0.02}_{-0.01}$ \\
2--60 keV  Flux\tablenotemark{c} (10$^{-10}$~erg~s$^{-1}$cm$^{-2}$) & 34.8$^{+3.6}_{-3.5}$\\   
\cutinhead{Gaussian Emission Line 1}
Energy (keV) & 8.6$^{+0.1}_{-0.2}$\\
Width\tablenotemark{e}, $\sigma$ (keV)  &  \nodata \\
2--60 keV  Flux\tablenotemark{c} (10$^{-10}$~erg~s$^{-1}$cm$^{-2}$) & 0.6$^{+0.2}_{-0.2}$\\   
\cutinhead{Gaussian Emission Line 2}
Energy (keV) & 14.2$^{+0.3}_{-0.3}$\\
Width, $\sigma$ (keV)  &   1.3$^{+0.3}_{-0.3}$  \\
2--60 keV  Flux\tablenotemark{c} (10$^{-10}$~erg~s$^{-1}$cm$^{-2}$) & 4.2$^{+0.9}_{-0.8}$\\   
\enddata
\tablenotetext{a}{All errors were extracted at the 1 $\sigma$ level.}
\tablenotetext{b}{In all spectral fits, the column density was held fixed at the value found by \citet{dv06b}.}
\tablenotetext{c}{All quoted fluxes are unabsorbed. }
\tablenotetext{d}{The blackbody radius was calculated assuming a distance to the source of 3.5~kpc \citep{dv06a}.}
\tablenotetext{e}{The width of this Gaussian line was not resolved, thus a Delta function was used smeared by the instrument response.}
\end{deluxetable}

\subsection{Pulse Profile Changes}
\label{sec:profiles}

\begin{figure*}
\centerline{\includegraphics[width=1.75\columnwidth]{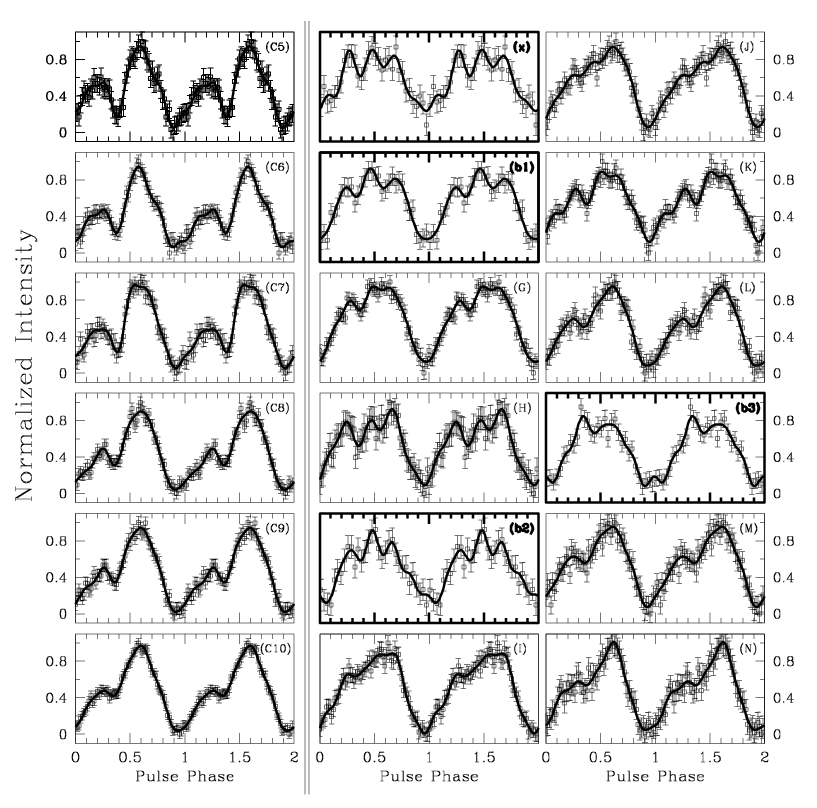}}
\caption
{ Pulse profile evolution of 4U~0142+61. Left-most column:  normalized average
  2$-$10~keV pulse profiles in the 6 years prior to the entry into the
  active phase in chronological order \citep[panels from ][]{dkg07}.
  Middle and right-most columns:  normalized,
  averaged 2$-$10~keV pulse profiles for each of the data groups
  marked by a letter in Figure~\ref{fig:observations}, after the entry
  into the active phase. The 4 plots with bold labels correspond to the
  observations marked with ``x'', ``b1'', ``b2'', and ``b3'' in
  Figure~~\ref{fig:observations}.
\label{fig:profiles}
}
\end{figure*}

\begin{figure*}
\centerline{\includegraphics[width=1.75\columnwidth]{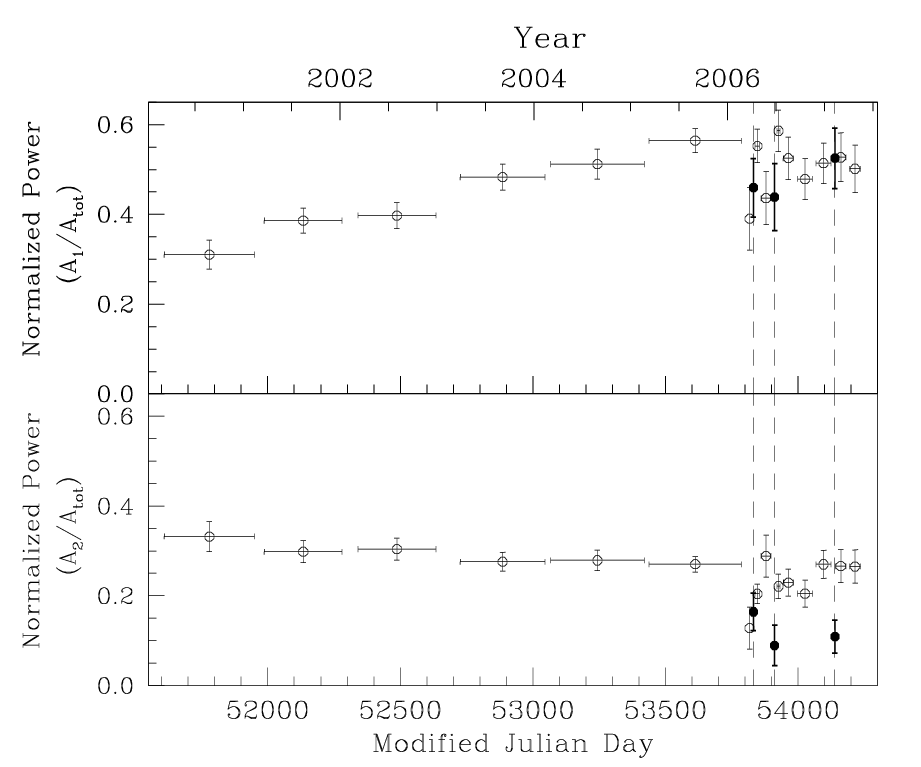}}
\caption{ Top: evolution of the power in the fundamental Fourier
  component of the pulse profile of 4U~0142+61 in the 2$-$10~keV
  band. Bottom: Evolution of the power in the second harmonic. In both
  panels, the points with large horizontal error bars are from
  \cite{dkg07}. The remaining points are obtained from groups of
  observations after the entry into the active phase.  The three bold
  points correspond to the observations with bursts. The three dashed
  lines correspond to burst epochs.
\label{fig:harmonics}
}
\end{figure*}

Many AXP outbursts are accompanied by significant pulse profile changes
\citep[e.g.][]{kgw+03}.  To search for these in 4U~0142+61,
for each observation, we generated 64-bin pulse profiles using the method described in
\cite{dkg07}. We then aligned the profiles with a high signal-to-noise
template using a cross-correlation procedure in the Fourier domain.
Then, for each group of observations in the active phase, we summed
the aligned profiles, extracted the DC component from the summed
profile, and scaled the resulting profile so that the value of the
highest bin is unity and the lowest bin is zero.

The resulting pulse profiles are shown in
Figure~\ref{fig:profiles} in chronological order. The panels in the left-most column are from
\cite{dkg07} and show pulse profiles
in the 6 years prior to the entry into the active phase. Notice the
slow change in the height of the dip between the peaks. The panels in
columns 2 and 3, marked with the corresponding letter in the top right
corner, show pulse profiles for each of the
data groups in the active phase that were marked with a letter in
Figure~\ref{fig:observations}. The 4 plots in bold correspond to the
observations marked with ``x'', ``b1'', ``b2'', and ``b3'' in
Figure~\ref{fig:observations}.

The pulse profile evolution can be described as follows: prior to the
active phase, features in the double peaked pulse profile were
evolving on a timescale of several years (see panels C6 to C10
corresponding to \xte\ Cycles 6 to 10). Then, in the first observation
of the active phase (panel x), the pulse profile became suddenly triple-peaked. 
It was also triple-peaked in the following observation (panel
b1), in which a burst occurred. It remained multi-peaked for the
following two groups of observations (panels G and H). Then, an
observation with multiple bursts occurred (panel b2). In that
observation, also having a triple-peaked pulse profile, the middle
peak was taller than the other two. Following this observation, the
pulse profile seemed to be slowly recovering back to its double-peaked
long-term shape (panels I, J, K, and L). Another burst observation
interrupted this evolution seven months later (panel b3). In that
observation, a large burst was detected. In the pulse profile of that
observation, the left-most peak was significantly taller than
usual. The event that caused this change had apparently no effect on
the following observation which occurred 2 days later: the profile
went back to being double-peaked (panels M and N). To summarize, the
pulse profile became multi-peaked at the beginning of the active
phase. Following the second observation with bursts, the profile
started to recover to its double-peaked shape. The evolution was only
temporarily interrupted for the duration of the third observation with
bursts. The behavior of the pulse profile in the 2$-$4 and
4$-$10~keV bands was similar.

Note that from Figure~\ref{fig:profiles} alone we can compare the
sizes of the peaks to each other, but we cannot track the evolution of
the heights of each peak separately. In order to do that, we must
scale the pulse profile of each group of observations by the average
pulsed flux of that group. This analysis is presented in
Section~\ref{sec:combo}.

We also performed an analysis of the Fourier components of the pulse
profiles. The results are shown in Figure~\ref{fig:harmonics}. The
variations in the power of the fundamental Fourier component are shown
in panel~1, and that of the second harmonic in panel~2.  
Note how the amplitude of the fundamental varied monotonically prior
to, but not during, the active phase.  Also note how the power in the
second harmonic was already back to its pre-active-phase level before
the last burst occurred.

\subsection{Burst Rotational Phases}
\label{sec:phases}

\begin{figure*}
\centerline{\includegraphics[width=1.75\columnwidth]{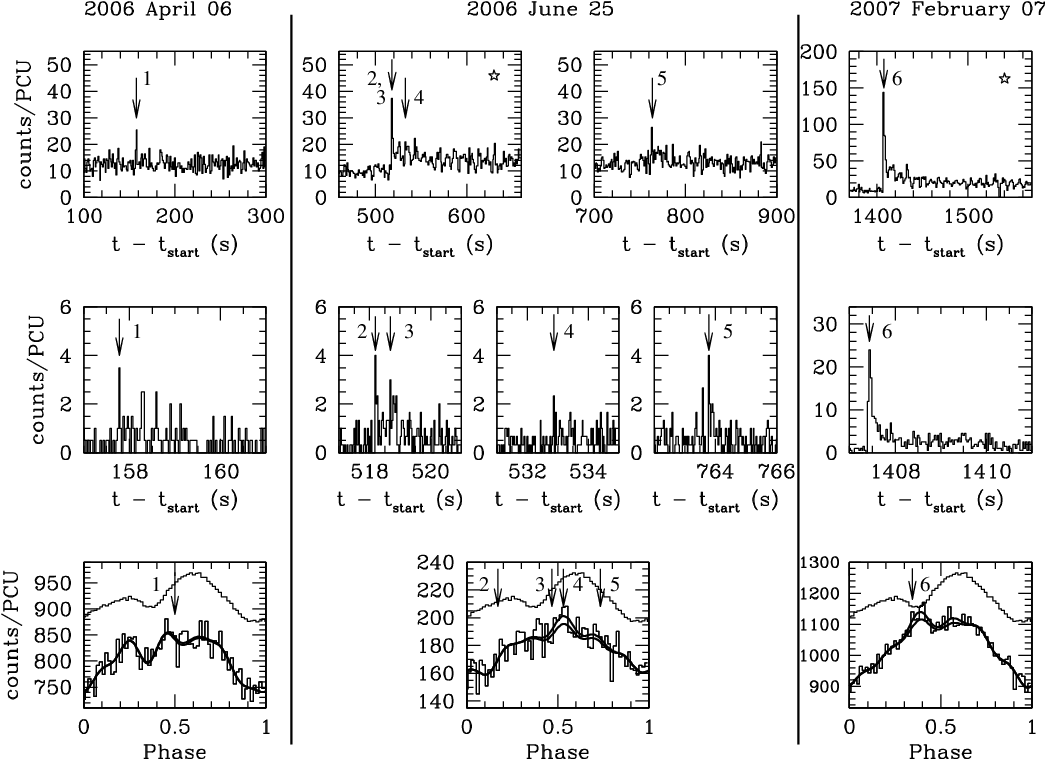}}
\caption
{ Phase analysis of the 6 detected bursts. Each column corresponds to
  an observation in which bursts were detected. Top row: 200-s segments
  of the time series containing the bursts in the 2$-$20~keV band. 
  The time resolution is
  1~s. Middle row: 4-s segments of the time series containing the
  bursts. The time resolution 31.25~ms. Bottom row: aligned folds of
  the burst observations shown below the scaled long-term average profile.  
  The folded counts are presented as
  histograms. Superposed on each histogram are two curves. The top
  curve is made of the 5 Fourier harmonics that best fit the
  histogram. The bottom curve is made of the best-fit 5 harmonics
  after the removal of the 4 seconds centered on each burst. The
  arrows indicate the phases of each burst.
\label{fig:phases}
}
\end{figure*}

An important factor in understanding the origin of
the bursts is the rotational phase at which they occur. The phases of
the bursts are shown in Figure~\ref{fig:phases}.  For each burst
observation we created at 31.25-ms time resolution lightcurve and
folded it using our timing solution (see \S~\ref{sec:timing}). We then
phase-aligned these folded profiles by cross-correlating them with the
long-term pulse profile template. Our phase-aligned folded profiles
are shown in Figure~\ref{fig:phases} (histograms in the last row).
Superposed on each folded profile are two curves. The top curve is
made of the 5 Fourier harmonics that best fit the histogram. The
bottom curve is made of the best-fit 5 harmonics after the removal of
the 4 seconds centered on each burst.  Note how the two curves in a
given plot are similar, demonstrating that the additional peaks in the
profiles are not due to the burst.  In fact, for the brightest (and
longest) burst, a fold of the last 1700 seconds of the data set shows
that by the time the observation ended, the pulse profile had not
recovered to its pre-burst shape.

The first three bursts occurred near the middle
``new'' peak in the profile.  Burst 5 occurred near the old tall peak
of the profile (see Fig.~\ref{fig:profiles}). The phase of burst 6
corresponds again to a new peak in the profile, this time located
where the previous small peak used to be.  The coincidences of several
bursts with new, transient profile features that are present even when
the actual burst data are removed are suggestive of lower-level
transient emission from the same physical location, with the burst
being the extreme of this emission's luminosity distribution.

\subsection{Pulsed Flux Analysis}
\label{sec:flux}
\subsubsection{Short-Term Pulsed Flux Analysis}
\label{sec:fluxshort}

\begin{figure*}
\centerline{\includegraphics[width=1.75\columnwidth]{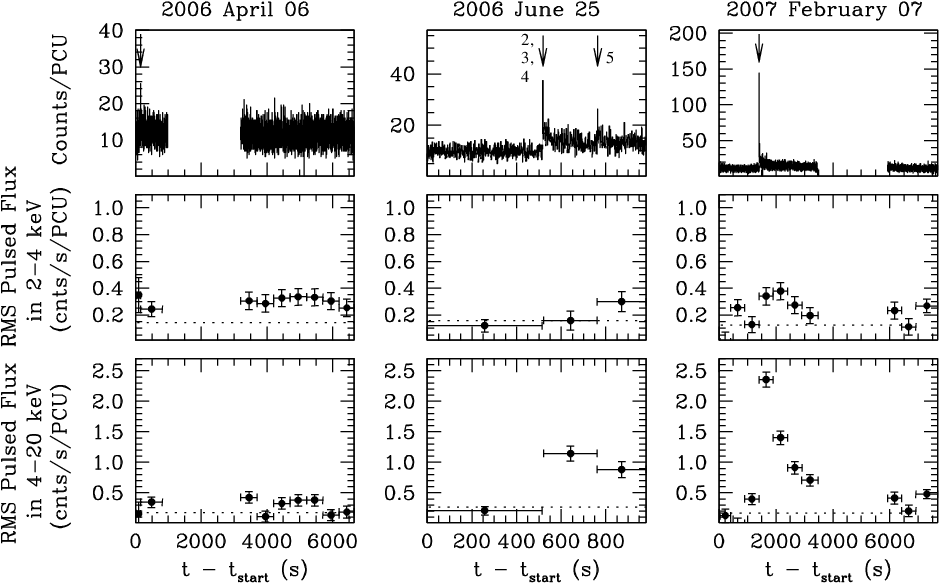}}
\caption{ RMS pulsed flux within the observations containing
bursts. Each column corresponds to one observation. In each column we
show, descending vertically, the 1-s resolution lightcurve with the
bursts indicated, the 2--4~keV RMS pulsed flux, and the 4--20~keV RMS
pulsed flux. The dotted line in each of the pulsed flux plots shows
the average of the pulsed fluxes obtained after segmenting and
analyzing the time series of the observation immediately prior to the
one shown.\label{fig:fluxshort}}
\end{figure*}

Previous AXP bursts are often accompanied by short-term pulsed
flux enhancements \citep[e.g.][]{gkw06}.  To search for these,
for each of the three observations containing bursts, we made two
time series in count rate per PCU, one for the 2$-$4~keV
band and the other for 4$-$20~keV. We included only photons
detected by PCUs that were on for the entire duration of the
observation. The time resolution was 31.25~ms. We removed the 4~s
centered on each burst from each time series. Then, we broke each time
series into segments of length $\sim$500~s. For each segment, we
calculated the pulsed flux using two different methods.

First, we calculated the RMS pulsed flux using
\begin{equation}
F_{\mathrm{RMS}} = 
{\sqrt{2 {\sum_{k=1}^{n}}
(({a_k}^2+{b_k}^2)-({\sigma_{a_k}}^2+{\sigma_{b_k}}^2))}},
\label{eq:f1}
\end{equation}
where $a_k$ is the $k^{\textrm{\small{th}}}$ even Fourier component
defined as $a_k$ = $\frac{1}{N} {\sum_{i=1}^{N}} {p_i} \cos {(2\pi k
i/N})$, ${\sigma_{a_k}}$ is the uncertainty of $a_k$, $b_k$ is the
$k^{\textrm{\small{th}}}$ odd Fourier component defined as $b_k$ =
$\frac{1}{N} {\sum_{i=1}^{N}} {p_i} \sin {(2\pi k i/N})$,
${\sigma_{b_k}}$ is the uncertainty of $b_k$, $i$ refers to the phase
bin, $N$ is the total number of phase bins, $p_i$ is the count rate in
the $i^{\textrm{\small{th}}}$ phase bin of the pulse profile, and $n$
is the maximum number of Fourier harmonics to be taken into account;
here $n$=5 for consistency with \cite{dkg07} and \cite{gdk+10}.

While least sensitive to noise compared to other pulsed flux
measurement methods, the RMS method returns a pulsed flux number that
is affected by pulse profile changes.  To confirm our pulsed flux
results, we also used an area-based estimator to calculate the pulsed
flux:
 \begin{equation}
 F_{\mathrm{Area}} = \frac{1}{N}{\sum_{i=1}^{N}} {p_i}   -  p_{\mathrm{min}},
  \label{eq:f2}
 \end{equation}
where $p_{\mathrm{min}}$ is the average count rate in the off-pulse
phase bins of the profile, determined by cross-correlating with a high
signal-to-noise template, and calculated in the Fourier domain after
truncating the Fourier series to 5 harmonics.  The results are shown
in Figure~\ref{fig:fluxshort} for $F_{\mathrm{RMS}}$
($F_{\mathrm{Area}}$ gives consistent results).  Note the significant
increase in the 4$-$20~keV pulsed flux in the 2006 June observation
following the cluster of bursts.  This increase is not present in
2$-$4~keV.  Also note the significant rise and subsequent decay of the
pulsed flux following the large 2007 February burst.  The pulsed flux
was sufficiently enhanced in these two observations that one can see
individual pulsations by eye in Figure~\ref{fig:phases} in the two
panels marked with a star, containing the raw burst lightcurves with
1-s time resolution.

\subsubsection{Long-Term Pulsed Flux Analysis}
\label{sec:fluxlong}

\begin{figure}
\centerline{\includegraphics[width=\columnwidth]{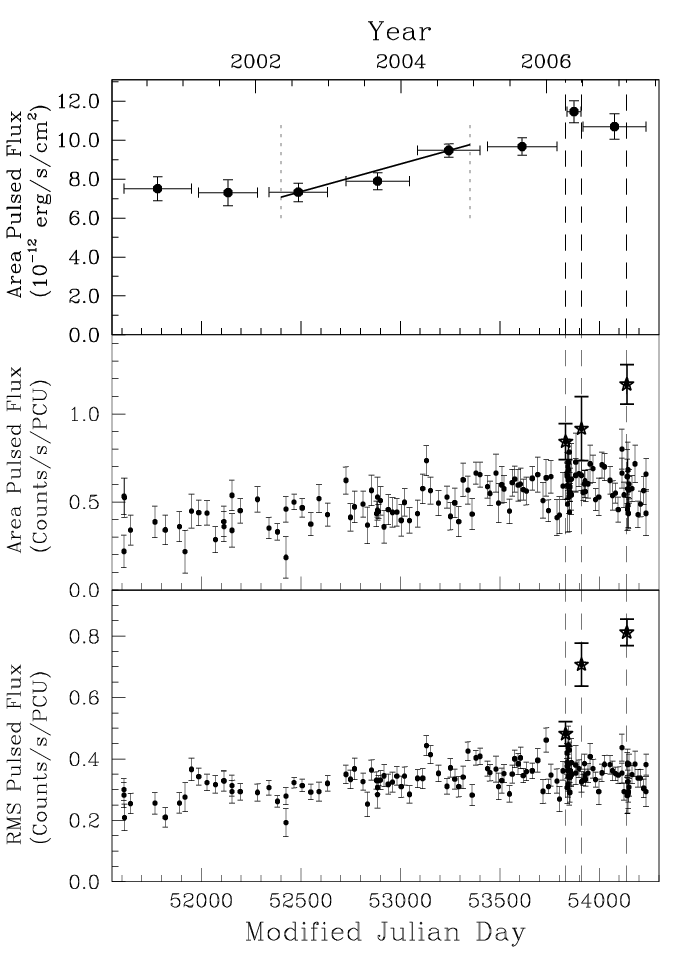}}
\caption { Long-term pulsed flux time series in 2$-$10~keV for
  \oft. Panel~1: area pulsed flux in erg~s$^{-1}$~cm$^{-2}$ for combined
  observations. The solid line marks the 29$\pm$8\% increase reported
  in~\cite{dkg07}. Panel~2: area pulsed flux in counts/s/PCU for
  individual observations. Panel~3: RMS pulsed flux in counts/s/PCU
  for individual observations. All panels: the dashed lines mark the
  burst epochs. The points marked with stars are the pulsed fluxes of
  the observations containing bursts.
\label{fig:fluxenergy}
}
\end{figure}

For each of the 94 analyzed observations, we created a pulse profile
(in units of count rate per PCU) using the same procedure as in
Section~\ref{sec:profiles}.  Then we calculated the pulsed flux for
each observation using Equations~\ref{eq:f1} and~\ref{eq:f2}. Data
from PCU~0 were omitted because the long-term trend in the pulsed
counts is not the same as that in the remaining PCUs, presumably due
to the loss of its propane layer.  Data from PCU~1 were omitted after
the loss of its propane layer as well, on MJD 54094.

We extracted pulsed fluxes in the 2$-$4 and 4$-$20~keV bands using
both the RMS and area pulsed flux method because of the numerous pulse
profile changes around the times of the bursts.
There is no evidence of a long-term change in the pulsed flux
associated with the bursts except, possibly, for a hint of an increase
in the 2$-$4~keV band roughly midway between the second and third
observations containing bursts. We also note that the 4$-$20~keV
pulsed flux of two of the observations containing bursts are
significantly larger than the long-term average. Removing the bursts
from these observations does not change this result.

We performed the same analysis for individual observations in 2$-$10~keV for
an extended period of time. This is shown in panels~2 and~3 of
Figure~\ref{fig:fluxenergy}. This plot is an update to Figure~10
of \cite{dkg07}.

In order to verify that trends seen in panels~2 and~3 of
Figure~\ref{fig:fluxenergy} are not an artifact of the evolution of
the response of the detector, we calculated the pulsed flux in
erg~s$^{-1}$~cm$^{-2}$ using a method that took the evolution of the
response into account.  The method is described in \cite{dkg07} and
takes into account spectral fits obtained from imaging data \citep[in
this case {\textit{XMM}} data, see][]{gdk+10} to convert counts to
energy for each combined set of observations. For this analysis, we
used data from all PCUs; however, data from PCU~1 were excluded after
the loss its propane layer.  Data from PCU~0 were included because the
response matrices used took into account the loss of its propane
layer.

The results are shown in panel~1 of Figure~\ref{fig:fluxenergy}. The
first 6 points, corresponding to {\emph{RXTE}} Cycles 6$-$10, are from
\cite{dkg07}.  The second to last point is obtained by combining all
observations that occurred between bursts~1 and ~2, but omitting
observations containing bursts.  The observations that we included
took place during the exponential recovery of the possible glitch (see
Section~\ref{sec:timing}). The last point in panel~1 was obtained by
combining the observations that occurred after burst 2, but again
omitting those containing bursts.  The observations we included took
place after the end of the exponential recovery of the possible glitch
(see Section~\ref{sec:timing}).

In the first of the two data points in the active phase, the pulsed
flux in erg~s$^{-1}$~cm$^{-2}$ is 18$\pm$8\% larger than in the
pre-active phase. This is consistent with the increase reported in
\cite{gdk+10} in the same energy range. A hint of this increase can be
seen in panel~2 although it appears less significant. This discrepancy
can be accounted for by the fact that the spectrum changed:
\cite{gdk+10} reported a temporary increase in the spectral hardness
in an {\emph{XMM}} observation of 4U~0142+61 immediately following the
bursts.

\subsubsection{Combined Pulse Morphology and Pulsed Flux Analysis}
\label{sec:combo}

\begin{figure}
\centerline{\includegraphics[width=0.82\columnwidth]{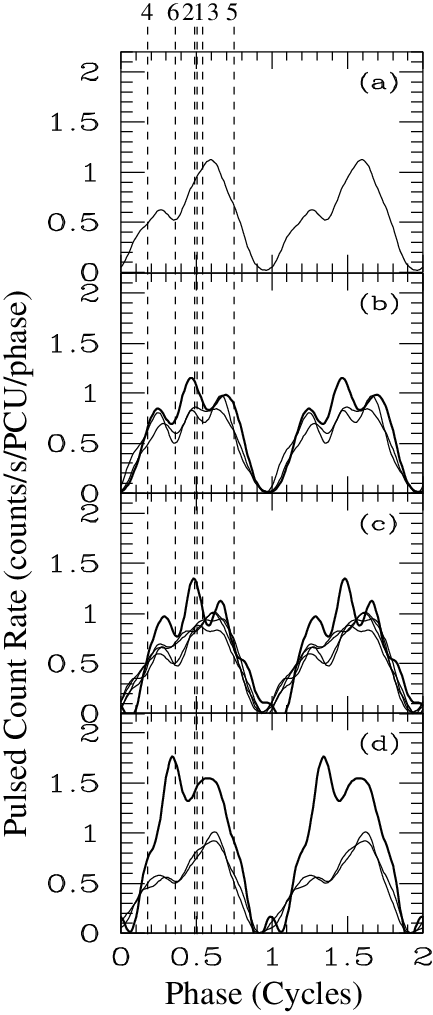}}
\caption
{ Average pulse profile per observation in the 2$-$10~keV band in several groups of
observations reconstructed from the first five Fourier components, and
scaled to return the appropriate pulsed flux. {\emph{(a)}} average
scaled pulse profile for observations in {\emph{RXTE}} Cycle~10, the
last {\emph{RXTE}} Cycle before the active phase. {\emph{(b)}} Bold
curve: scaled pulse profile of the observation containing
burst~1. Thin curves: average scaled pulse profile for each of the
groups of observations that followed burst~1 (groups~G and~H in
Figure~\ref{fig:observations}). {\emph{(c)}} Bold curve: scaled pulse
profile of the observation containing bursts~2, 3, 4, and~5. Thin
curves: average scaled pulse profile for each of the groups of
observations that followed the bursts (groups~I, J, K, and~L in
Figure~\ref{fig:observations}). {\emph{(d)}} Bold curve: scaled pulse
profile of the observation containing burst~6. Thin curves: average
scaled pulse profile for each of the groups of observations that
followed burst~6 (groups~M, and~N from Figure~\ref{fig:observations}).
The vertical dotted lines indicate the phases of each burst.
\label{fig:scaled}
}
\end{figure}

In Section~\ref{sec:profiles}, we calculated the Fourier components of the
aligned average pulse profiles. This gave us the relative amplitude of the
pulse profile harmonics in each group of observations marked with a letter
in Figure~\ref{fig:observations}. In Section~\ref{sec:fluxlong}, we
calculated the pulsed flux for every observation. Here, we compute a
weighted average of the pulsed flux for each group of observations using the
flux points calculated in Section~\ref{sec:fluxlong}. We then reconstruct
the profiles for each of the groups from the first five Fourier components
(not including the DC), scale them by the average RMS pulsed flux for that
group, and add the necessary offset for the lowest point on each curve to be
zero. This means that the resulting scaled profiles return the correct
pulsed flux.  The advantage of this analysis is that we can now trace the
evolution of each of the peaks independently.

The results are presented in Figure~\ref{fig:scaled}. In panel (a), we show
the average scaled pulse profile per observation for observations in the
year preceding the active phase. The profile is double-peaked. In panel (b),
we show in bold the scaled profile for the observation containing burst~1.
We also show the scaled profiles in each of the groups of observations that
followed the burst. The profiles are triple-peaked. In panel (c), we show in
bold the scaled profile for the observation containing bursts~2, 3, 4,
and~5. We also show the scaled profiles in each of the groups of
observations that followed the bursts. Note how the increase in the pulsed
flux in the observation containing the bursts is not only a consequence of
the appearance of the new peak, but a result of the increase in
size of all three peaks. Also note the evolution of the pulse profile back
to being double-peaked. In panel (d), we show in bold the scaled profile for
the observation containing burst~6. We also show the scaled profiles in each
of the groups of observations that followed the burst. Note again how the
pulsed flux increase is due to both peaks increasing in size.

\subsection{Timing Analysis}
\label{sec:timing}

\begin{figure*}
\centerline{\includegraphics[width=1.75\columnwidth]{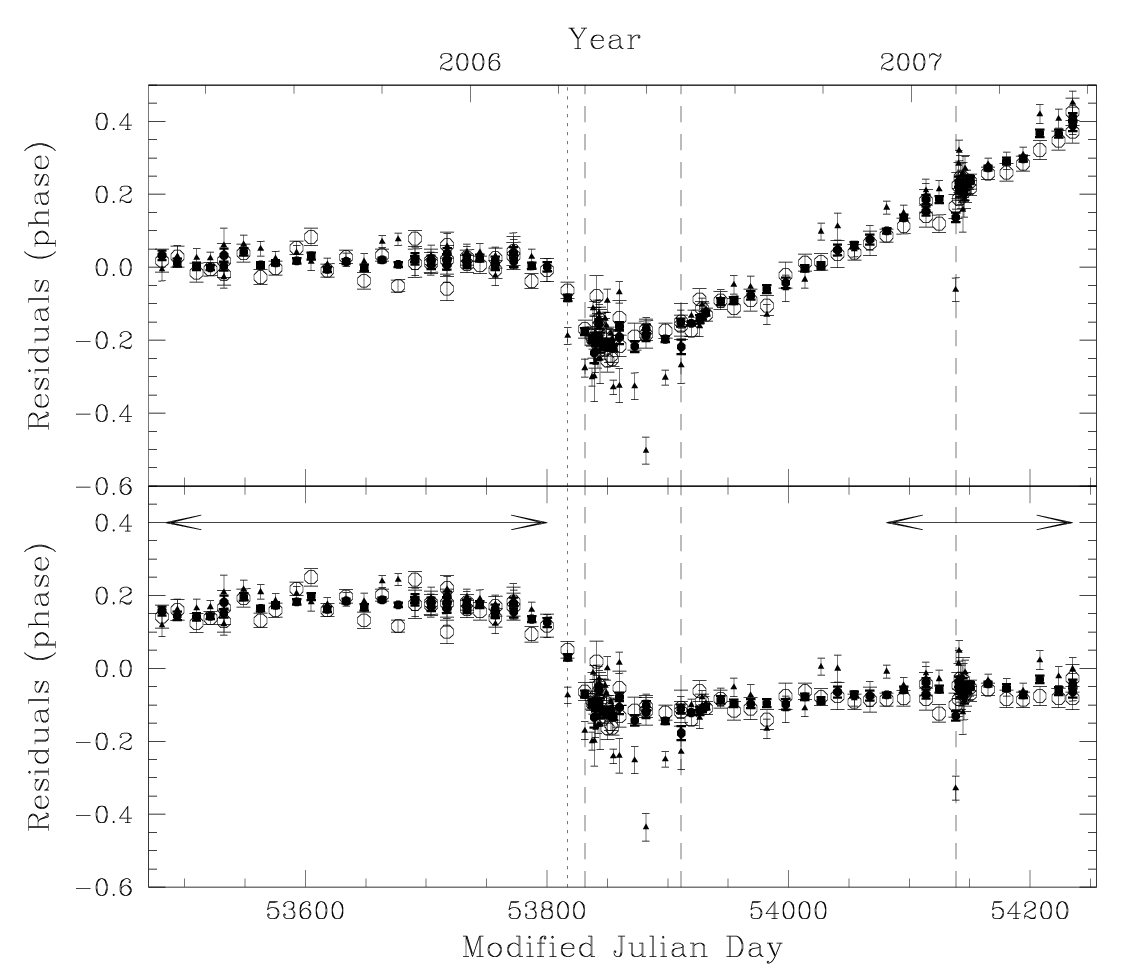}}
\caption {Top: Timing residuals for all three sets of TOAs. Bold
  circles: timing residuals obtained using the first set of TOAs
  obtained by cross correlation in the Fourier domain. Empty circles:
  residuals obtained using the second set of TOAs obtained by aligning
  the off-pulse regions of the pulse profiles. Triangles: residuals
  obtained using the third set of TOAs obtained by aligning the
  tallest peak of the pulse profiles. The linear ephemeris used to
  produce all three sets of residuals is that which best fit the first
  set of TOAs in the pre-active phase region.  Bottom: Timing
  residuals for the same three sets of TOAs. The linear ephemeris used
  is that which best fit the first set of TOAs exclusively in regions
  indicated with the horizontal arrows with an arbitrary phase jump
  in the gap between the arrows. The phase residuals are relative to the linear ephemeris (ignoring the phase jump) and vertically offset such that the average phase residual is zero.
  Both panels: the
dotted line indicates the epoch of the first observation of the active
phase. The dashed lines indicate the burst epochs.
\label{fig:toas} }
\end{figure*}

\begin{figure}
\centerline{\includegraphics[width=\columnwidth]{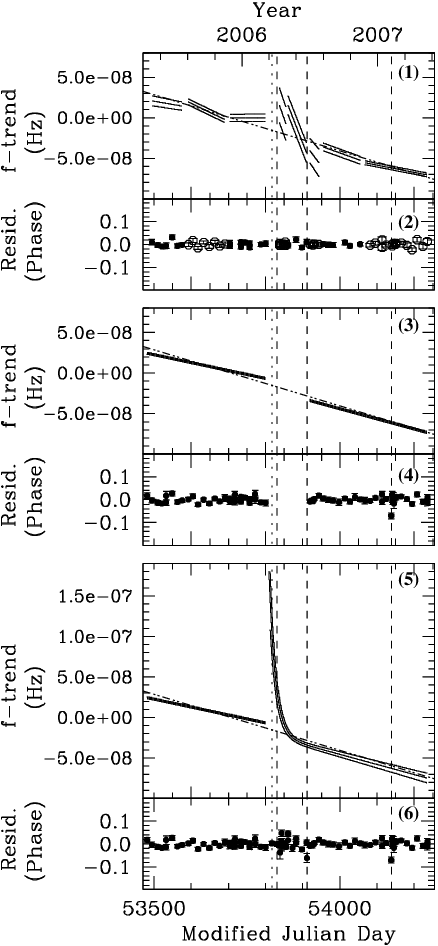}}
\caption{Panel~1: Frequency vs. time plot of local linear ephemerides
with uncertainties. The linear trend corresponding to the local
ephemeris directly before the active phase is subtracted from all
local ephemerides.  Panel~2: Timing residuals after subtracting the
ephemerides shown in panel~1.  Panel~3: Frequency vs. time plot (with
uncertainties) of the best-fit linear ephemerides in the pre-active
phase region and in the post burst~2 region. The same trend as in
panel~1 was subtracted.  Panel~4: Timing residuals after subtracting
the ephemerides shown in panel~3.  Panel~5: Frequency vs. time plot
(with uncertainties) of the best-fit long-term ephemeris which
includes a glitch at MJD 53809 followed by a fast exponential
recovery. The same trend as in panel~1 was subtracted.  Panel~6:
Timing residuals after subtracting the ephemeris shown in panel~5.
All panels: the
dotted line indicates the epoch of the first observation of the active
phase. The dashed lines indicate the burst epochs.
\label{fig:timing}
}
\end{figure}

Many AXP outbursts and active phases are accompanied by interesting
timing anomalies \citep[e.g.][]{kgw+03,wkt+04,icd+07,dkg09}.  Next
we consider the timing behavior of 4U~0142+61, which, as we show,
also exhibits interesting evolution at the start of the active phase.

For all our \xte\ observations of 4U~0142+61 we created 31.25-ms time
resolution lightcurves, including only the events in the energy range
2.5$-$9~keV to maximize the signal-to-noise ratio of the pulse.  Each
binned time series was epoch-folded using an ephemeris determined
iteratively by maintaining phase coherence; see below.  Resulting
pulse profiles, with 64 phase bins, were cross-correlated in the
Fourier domain with a high signal-to-noise template created by adding
phase-aligned profiles from all observations. The cross-correlation
returned an average pulse time of arrival (TOA) for each observation
corresponding to a fixed pulse phase. The pulse phase $\phi$ at any
time $t$ can be expressed as a Taylor expansion,

\begin{eqnarray}
\phi(t) & = &
\phi_{0}(t_{0})+\nu_{0}(t-t_{0})+
\frac{1}{2}\dot{\nu_{0}}(t-t_{0})^{2} \nonumber \\
& & +\frac{1}{6}\ddot{\nu_{0}}(t-t_{0})^{3}+{\ldots},
\end{eqnarray}

where $\nu$~$\equiv$~1/$P$ is the pulse frequency,
$\dot{\nu}$~$\equiv$~$d\nu$/$dt$, etc$.$, and subscript `0'' denotes a
parameter evaluated at the reference epoch $t=t_0$. To obtain ephemerides,
the TOAs were fitted to the above polynomial using the pulsar timing
software package TEMPO\footnote{See
http://www.atnf.csiro.au/research/pulsar/tempo.}.

As explained above, our first set of TOAs was obtained by aligning the
folded observations with a template profile using a cross-correlation
procedure. In order to determine to what extent this set of TOAs was
affected by the pulse profile changes that took place in the active phase,
we generated two additional sets of TOAs (sets~2 and 3).

For set~2, we made the assumption that the location of the trough of the pulsations,
determined by finding the minimum in the pulse profile after 
smoothing, is not affected by the pulse
profile changes. We then generated TOAs by aligning the pulsation trough of 
the folded observations with that of the  template and
extracting phase differences. 
The last set of TOAs (Set~3) was
obtained after aligning the tallest peak of each smoothed pulse profile with
that of the long-term template. 
All three sets of TOAs are provided as a machine readable table in
the electronic edition. A portion of the table starting from the
active phase is provided in Table~\ref{ta:TOAs}.

\begin{deluxetable*}{cccccccccccccc}
\tablewidth{2.0\columnwidth}
\tablecaption{A Sample of our \xte\ timing observation log for \oft. The full version of this table is provided in the electronic edition of the journal.\label{ta:TOAs}}
\tablehead{
\colhead{}    & \colhead{}        & \colhead{}         & \colhead{}       & \multicolumn{2}{c}{\underline{\ \ \ \ \ Set 1\ \ \ \ \ }} &  \multicolumn{2}{c}{\underline{\ \ \ \ \ Set 2\ \ \ \ \ }} & \multicolumn{2}{c}{\underline{\ \ \ \ \ Set 3\ \ \ \ \ }} &  \\
\colhead{TOA} & \colhead{Obs. ID\tablenotemark{b}} & \colhead{Exp.\tablenotemark{c}} & \colhead{Active} & \colhead{TOA\tablenotemark{e}} & \colhead{$\sigma_{\mathrm{TOA}}$\tablenotemark{f}} & \colhead{TOA\tablenotemark{e}} & \colhead{$\sigma_{\mathrm{TOA}}$\tablenotemark{f}} & \colhead{TOA\tablenotemark{e}} & \colhead{$\sigma_{\mathrm{TOA}}$\tablenotemark{f}} & \colhead{PT}  \\
\colhead{Numb.\tablenotemark{a}} & \colhead{} & \colhead{}  & \colhead{PCUs\tablenotemark{d}} & \colhead{} & \colhead{}    & \colhead{} & \colhead{}    & \colhead{} & \colhead{}  & \colhead{Analysis?\tablenotemark{g}}  \\
\colhead{} & \colhead{} & \colhead{(ks)}  & \colhead{} & \colhead{(MJD)} & \colhead{(s)} & \colhead{(MJD)} & \colhead{(s)}    & \colhead{(MJD)} & \colhead{(s)}  & \colhead{} }
\startdata
\multicolumn{13}{c}{$\cdots\cdots\cdots\cdots$}\\
33 & 92006-05-02-00 & 4.9 &   3  & 53817.1615364  & 0.08 & 53817.1616238 &  0.20 & 53817.1615403 & 0.20 & Yes   \\
34 & 92006-05-03-00 & 4.5 &   3  & 53831.3358686  & 0.09 & 53831.3358558 &  0.21 & 53831.3358711 & 0.21 & Yes   \\
\multicolumn{13}{c}{$\cdots\cdots\cdots\cdots$}\\
58 & 92006-05-09-01 & 1.0 &   3  & 53911.0530210 & 0.17 &  53911.0531137 &  0.43 & 53911.0530298 & 0.43 & Yes  \\
\multicolumn{13}{c}{$\cdots\cdots\cdots\cdots$}\\
78 & 92006-05-25-00 & 5.3 &   2  & 54138.4487229 & 0.11 &  54138.4487002 &  0.29 & 54138.4487277 & 0.29 & No  \\
\multicolumn{13}{c}{$\cdots\cdots\cdots\cdots$}\\
\enddata
\tablenotetext{a}{TOA number. TOA number 33 is the first observation in the active phase. TOAs number 34, 58, and 78 represent the first, second and third observation containing bursts.}
\tablenotetext{b}{\xte\ observation identification number.}\tablenotetext{c}{The effective number of PCUs on during the observation.}
\tablenotetext{d}{Total exposure.}\tablenotetext{e}{Barycentered pulse TOA. Set corresponds to the TOAs obtained by cross-correlating the folded pulse profiles with a high signal-to-noise template. Sets  2 and 3 are   similar to Set 1, however, we constrained our cross-correlation to the trough and peak of the pulsations, respectively. See \S~\ref{sec:timing} for details.}\tablenotetext{f}{Uncertainty on TOA for Set 1, 2,and 3.}\tablenotetext{g}{Whether the TOA was included in the partial timing analysis. Note: All TOAs were included in the long-term analysis.}
\end{deluxetable*}

Timing residuals for all three sets of TOAs are shown in
Figure~\ref{fig:toas} using two different ephemerides. 
The linear ephemeris used in the top panel
is the ephemeris that best fit the first set of TOAs prior to the
active phase.  
The linear
ephemeris used in the bottom panel is that obtained by fitting the first
set of TOAs in the regions indicated by the horizontal arrows, with a
phase jump in between.  
The same long-term trend is
present in the first two sets of timing residuals, with more scatter
in the second set. The difference in phase between each bold circle
and the corresponding empty circle represents our uncertainty in
determining a fiducial point on the pulsar. In the third set of
residuals, the outlier points represent the observations where the
largest peak was no longer the right-most peak. Apart from the
outliers, the same trend seen in the other two sets of residuals is
present in the third set.

From here on, we assume that the presence of the same trend in our
three sets of residuals is an indication that the TOAs in the first
set were not significantly affected by the pulse profile changes. We
therefore have used the first set of TOAs (one for each of the 94 observations) in the remainder of this
Section.

Using the above assumption, the results of the timing analysis are
shown in Figure~\ref{fig:timing}. The results of a segmented timing
analysis are in panels~1 and ~2. The results of a long-term timing
analysis are in panels~3, 4, 5, and~6 of the same Figure.

For the segmented timing analysis, we used the 70 TOAs that had the
smallest uncertainties (indicated in the online version of
Table~\ref{ta:TOAs}) and omitted the TOAs of the 3 burst observations
as well as the TOA of the first observation in the active phase. We
then divided the data into segments of similar pulse profiles. For
each data segment we found a linear ephemeris using TEMPO\footnote{For
  the segmented timing analysis we ran TEMPO in mode~1.}. We plotted
the resulting ephemerides in panel~1 of Figure~\ref{fig:timing} with
the uncertainties returned by TEMPO. The timing residuals are shown in
panel~2. Notice how the slope between bursts~1 and ~2 is more negative
than the long-term average.

We then fit a linear trend through all the pre-burst observations, and
another linear ephemeris through the post-burst observations after the
pulse profile had started to return to the double-peaked shape. We did
not include the TOA corresponding to the last observation containing
bursts. The results are plotted in panel~3 of Figure~\ref{fig:timing}
with uncertainties, with the residuals in panel~4. Notice the
difference in the slope in the two regions. In particular,
extrapolating the two ephemerides to a point between their times of
validity makes it seem like a sudden spin down, i.e.  an
``anti-glitch'' occurred.

Finally, we included all the TOAs and did one global fit. In order to
provide a good fit to the TOAs at the onset of the active phase, we
had to assume that a glitch occurred on MJD 53809, with the glitch
model consisting of a permanent change in $\nu$ and $\dot{\nu}$ and a
frequency change $\Delta\nu_d$ that recovered exponentially on a timescale
$\tau_d$, i.e.,

\begin{equation}
\nu = \nu_0(t) + \Delta\nu + \Delta\nu_{d}e^{-(t-t_g)/{\tau_d}} +
\Delta\dot{\nu}~(t-t_g),
\label{eq:glitch}
\end{equation}

where $\nu_0(t)$ is the frequency evolution pre-glitch given by
$\nu_0(t)=\nu(t_0)+\dot{\nu}(t-t_0)$, $\Delta \nu$ is a instantaneous
unrecovered frequency jump, $\Delta \nu_d$ is the frequency increase
that decays exponentially on a time scale $\tau_d$, $t_g$ is the
glitch epoch determined by setting the phase jump to zero, and $\Delta
\dot{\nu}$ is the post-glitch change in the long-term frequency
derivative. The values of the fit parameters are listed in
Table~\ref{tab:timing}.

From Table~\ref{tab:timing}, we can see that the total sudden change
in frequency which happened at the onset of the active phase ($\Delta
\nu_{tot}$ = $\Delta \nu$ + $\Delta \nu_d$) is positive, and 
$\Delta \nu_d$ decayed exponentially. At the end of
the decay, the net effect of the recovered glitch was a negative
$\Delta \nu$. That is, the data suggest that after $\nu_d$ had
decayed, there remained net spin-{\emph{down}} relative
to the undisturbed ephemeris from before the active phase.

The fit parameter $\Delta \nu_d$, from which we conclude a sudden
spin-up, is controlled primarily by the first few TOAs in the active
phase. It is therefore possible that it is affected by pulse profile
changes.
However, as shown in Figure~\ref{fig:toas}, the TOAs in the active
phase were not significantly contaminated by pulse profile
changes. Also, the segmented analysis shown in Figure~\ref{fig:timing}
clearly indicates an initial spin-up.

In contrast to $\Delta \nu_d$, the fit parameter $\Delta \nu$, from
which we conclude a net spin {\emph{down}}, is primarily
controlled by the TOAs outside of the active phase, which were certainly not
affected by pulse profile changes (Figure~\ref{fig:timing},
panel~3).

We therefore conclude that the pulsar likely suffered a spin-up glitch
near or at the start of the radiatively active phase, but that the glitch
`over-recovered' such that long-term, its net effect is a spin-{\it down}
of the pulsar.

\begin{deluxetable}{lc}
\tablewidth{\columnwidth}
\tablecaption
{
Spin and Glitch Parameters for 4U~0142+61\tablenotemark{a}
\label{tab:timing}
}
\tablehead
{
\colhead{Parameter} &
\colhead {Value}
}
\startdata
MJD start & 53481.268\\
MJD end & 54235.319\\
TOAs & 93\tablenotemark{b}\\
$\nu$ (Hz) & 0.1150920955(12)\\
$\dot{\nu}$ (10$^{-14}$ Hz s$^{-1}$) &  $-$2.661(9)\\
Epoch (MJD) & 53809.185840\\
Glitch Epoch (MJD) & 53809.185840\\
$\Delta\nu$ (Hz) & $-$1.27(17)$\times$10$^{-8}$\\
$\Delta\dot{\nu}$ (Hz s$^{-1}$) & $-$3.1(1.2)$\times$10$^{-16}$\\
$\Delta \nu _{d}$ (Hz) & 2.0(4)$\times$10$^{-7}$\\
$t_d$ (days) & 17.0(1.7)\\
RMS residual (phase) & 0.0168
\enddata
\tablenotetext{a} {Numbers in parentheses are TEMPO-reported 1$\sigma$
uncertainties. }
\tablenotetext{b} {A single TOA was omitted due to the very poor signal to
noise ratio in the corresponding observation.}
\end{deluxetable}

\section{Discussion}
\label{sec:discussion}

In this paper, we have described the timing, pulse profile, and pulsed
flux behavior of \oft, during its 2006-2007 active phase, the first
such episode yet studied from this source.  Specifically we have shown
that in addition to a sudden departure from a slow, systematic
evolution of the source's pulsed flux and pulse profile, this AXP also
suffered a significant timing event that is best described as a sudden
spin-up glitch, followed by a large decay of the frequency jump such
that the net effect was a slow-{\it down} with $\Delta\nu/\nu \simeq
-8\times 10^{-8}$.  Interestingly, the pulsed X-ray flux showed no
significant change apart from immediately following bursts, in
contrast to other AXP radiative outbursts
\citep[e.g.][]{kgw+03,tgd+08,ims+04,mgc+07}.  Further, we have
reported on six SGR-like bursts from the source that occurred during
this period, one of which was notable for its unusual spectrum, which
was poorly fit by an continuum model.

\subsection{The SGR-Like Bursts from \oft}

Five of the six bursts reported here for \oft\ had
fast-rise-slow-decay profiles, with tails much longer than the rise
times.  These morphologies are similar to the class of AXP bursts
labeled as ``Type B'' by \citet{wkg+05}.  Type B AXP bursts have also
been characterized by arriving preferentially at pulse maximum, and
are seen more often in AXPs compared with SGRs \citep{wkg+05,gkw06}.
They were suggested as being due to a sudden rearrangement of magnetic
field lines anchored in the crust following a crustal fracture
\citep{td95}, as opposed to reconnection events in the upper
magnetosphere \citep[e.g.][]{lyu02}, argued as more likely for the
shorter, symmetric ``Type A'' bursts that show no preference for pulse
maximum.  However, it is interesting that in the \oft\ bursts, in
spite of their B-type morphology, no clear preference for arrival at
or near pulse maximum was seen (see Table ~\ref{tab:bursts}).  This
may blur somewhat the distinction between the putative types, although
the morphological distinction remains clear.

It is also notable that the vast majority of bursts seen from AXPs
had long tails following fast rises.  An exception
 is \tfn\ for which, during its 2002 outburst, the minority
(roughly one dozen out of 80) of the bursts seen were of this form,
the majority being of Type A, similar to those classically seen in
SGRs, and suggested by \citet{wkg+05} to be magnetospheric.  A
possible hint regarding the origin of the different types may lie in
that \tfn\ was mid-outburst when its bursts were observed, whereas the
other sources' bursts all occurred in the days/months following the
commencement of an active period, presumably as the pulsar recovered
from a major event, rather than mid-event.

\subsubsection{Spectral Features}

\begin{figure}
\centerline{\includegraphics[width=\columnwidth]{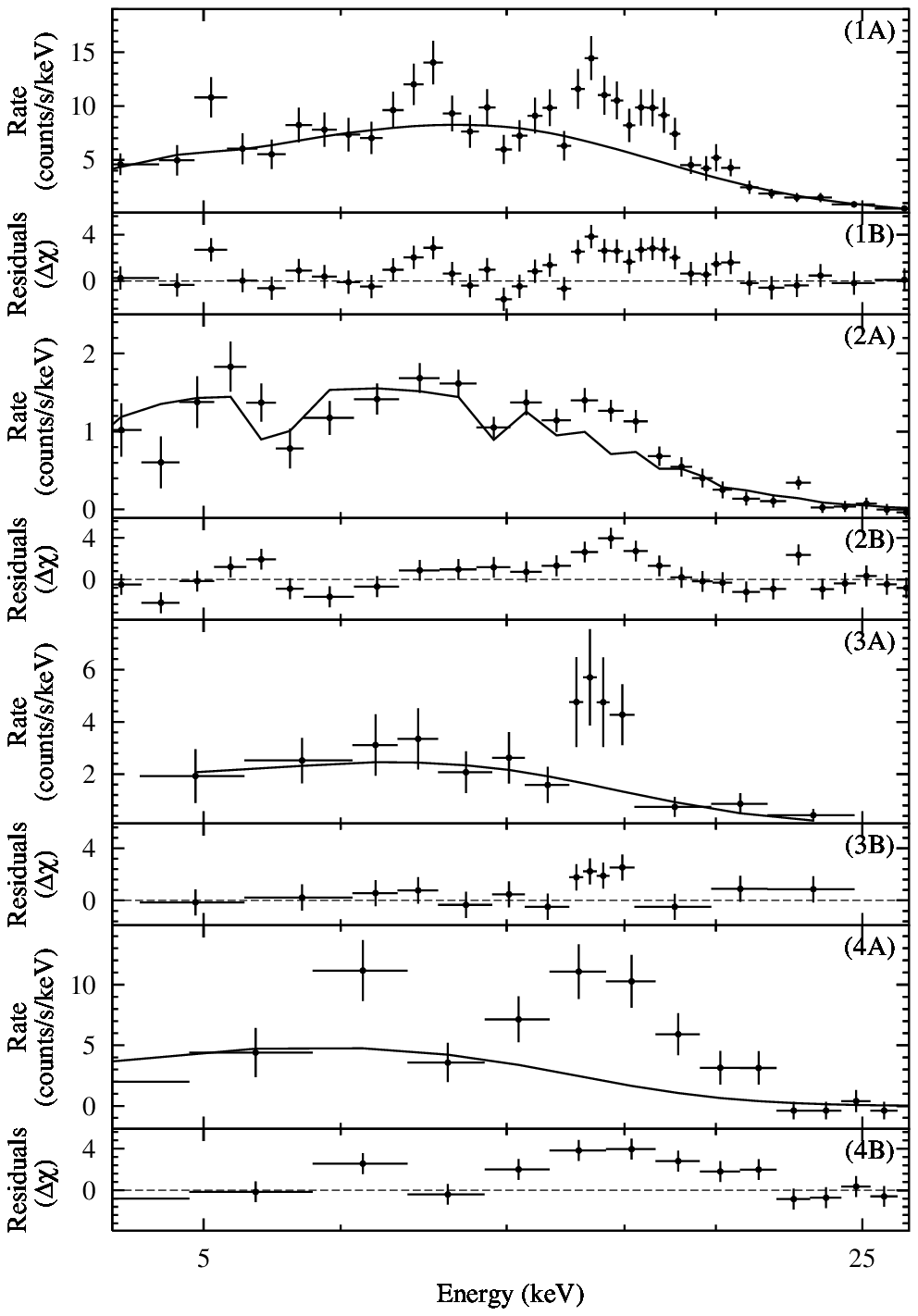}}
\caption{Burst spectra of all observed AXP bursts with significant
emission lines, as seen by \xte. Panel 1A: Burst spectrum of \oft\
burst 6. The solid line indicates the continuum (blackbody) component
of the best-fit model folded through \xte's instrumental response. Panel 1B: Residuals, expressed in terms of the
change in $\chi^2$, after subtracting only the continuum component of
the best-fit model. Panels 2A and 2B: Same but for burst 4 of \ett\
\citep[see][]{wkg+05}. Panels 3A and 3B: Same, but for burst 3 of
\tfe\ \citep[see][]{gkw06}. Panels 4A and 4B: Same, but for burst 1 of
\tfe\ \citep[see][]{gkw02}.
  \label{fig:manysources}}
\end{figure}

The complicated spectrum of burst 6 is puzzling.  The most broad
feature at $\sim$14 keV is particularly interesting. Emission features
at similar energies were observed from two out of the three bursts
from \tfe\ \citep{gkw02,gkw06} and in one out of the four bursts from
\ett \citep{wkg+05}. We have reanalyzed all these burst spectra
consistently as for \oft. In Figure~\ref{fig:manysources} we plot the
spectra of all AXP bursts with likely emission lines in their
spectra. Notice that all have broad features that occur between 13 and
14~keV. There is also evidence for a narrow feature at $\sim$8~keV in the
\oft\ burst; there may be hints of additional features in some of the
other burst spectra as well.

If the features are interpreted as of a cyclotron origin, it is
puzzling that three sources with spin-down-inferred magnetic fields
that span a range of a factor of three produce features at similar
energies.  The similarity in energy suggests a mechanism that is not
very sensitive to the exact value of the magnetic field.  Moreover, it
is unclear why such features are seen only in some bursts.  Cyclotron
features have been observed in the persistent spectra of
accretion-powered X-ray pulsars \citep{chr+02}.  Unlike the features
observed here, these features can be well fit by a Gaussian absorption
model (see \S~\ref{sec:burstspectral}). In the spectra of some
accretion-powered X-ray pulsars there is also evidence for higher
harmonics. The features observed in the \oft\ burst were not
consistent with being harmonically related, even after taking into
account the relativistic effects on the spacing \citep[see][]{hd91,ah99}.

Thus far, all AXP burst spectral features have been observed with
\rxte.  Their absence from many other \rxte-observed bursts, together
with the absence of any response feature in the PCA near 13-14 keV, is
supporting evidence for their veracity.  Nevertheless, it is important
to confirm these features with other instruments.  Observations of
AXPs during active phases using the {\it Swift} XRT or the large-area
LAXPC aboard the planned {\it Astrosat} mission would be particularly
useful in this regard.

\subsection{The Net Spin-Down Event in \oft}

The timing glitch reported on in \S\ref{sec:timing} had recovery
fraction, defined as $Q\equiv \Delta\nu_d/(\Delta\nu_d + \Delta\nu)$,
equal to $~1.07 \pm 0.02$.   $Q>1$ implies that the net frequency
change after the transients have decayed is negative (see panel~3 of
Figure~\ref{fig:timing} and Table~\ref{tab:timing}).  Indeed, the net
$\Delta\nu/\nu$ observed is $-1.1 \pm 0.1 \times 10^{-7}$.  Such a $Q>1$
has not previously been reported in any AXP but has been seen following
magnetar-like radiative behavior in one high-$B$ rotation-powered
pulsar \citep{lkg10}.
Note that \citet{mks05} reported a candidate glitch in 1999 from this
source. If such a glitch occurred, based on the ephemerides reported
in \cite{dkg07}, the fractional frequency change, after any recovery,
would have been +1.9(2)$\times 10^{-7}$$-$8.6(2)$\times 10^{-7}$
depending on the glitch epoch, but manifestly positive.

In the standard model for glitches in rotation-powered pulsars, the
neutron-star crust contains superfluid neutrons that rotate faster
than the bulk of the surrounding matter \citep[see,
e.g.,][]{ai75,aaps82,aaps84a,ap93}.  This is argued to be a result of
the fact that the external magnetic torque acts on the crust and
coupled core components only, with the uncoupled superfluid
unaffected.  The superfluid's angular momentum resides in quantized
vortex lines whose density is proportional to angular velocity. The
vortex lines are suggested to be pinned to crustal nuclei, and suffer
strong forces due to the angular velocity lag that builds between
crust and superfluid.  A glitch in this picture is a sudden unpinning
and outward motion of vortex lines, with a transfer of angular
momentum from superfluid to crust.  In a magnetar, unpinning may occur
due to the strong internal magnetic field as it deforms the crust
plastically or cracks it violently \citep{td96a}.

In the \oft\ event we describe in \S\ref{sec:timing}, if the standard
glitch model is roughly correct and applies here, then some regions of
the stellar superfluid were originally spinning {\it slower} than the
crust.  Then the transient increase in frequency would be a result of
transfer of angular momentum first from a more rapidly rotating
region, with a subsequent angular momentum drain from the crust to a
more slowly rotating region.  As argued by \citet{tdw+00}, regions of
slower-rotating superfluid can occur in magnetars, because the
superfluid vortex motions are governed not be spin-down-related
forces, but by advection across the stellar surface by the deforming
crust.  Those authors show that the number of vortex lines per unit
surface area of crust can increase or decrease depending on the
crustal motion relative to the stellar rotation axis.  They invoke
this possibility to explain a possible ``anti-glitch'' seen in
SGR~1900+14.  However, that event was orders of magnitude larger than
what we have observed in \oft, with $\Delta \nu/\nu \sim 10^{-4}$
\citep{tdw+00}, and also was likely associated with the extremely
energetic flare seen on 1998 August 27 \citep{mca+99}.

Independent evidence for slow crustal deformations in \oft\ that could
result in the regions of slower-rotating superfluid required to
explain the spin-down may come from the long-term flux and pulse
profile evolution previously reported for this source by \citet{dkg07}
and shown again here in Figures 5 and 6.  Although such variations
could also be magnetospheric in nature, possibly due to twisting of
the magnetic field \citep[e..g.][]{tlk02,bt07}, problems with this
interpretation exist, as discussed by \citet{dkg07}.  For example, if
magnetospheric, it is puzzling that most if not all of the changes in
pulse profile are seen below 4 keV, with none above 6 keV.  On the
other hand, slow evolution of the surface emission
\citep[e.g.][]{og07}, hence structure, is perhaps more consistent with
low-energy evolution.  It remains to be seen if detailed modeling can
reproduce the sort of slow pulse profile changes we have observed
(Fig. 5), quite apart from the sudden appearance of new peaks, hence
greater harmonic structure, near bursts.  We note that the latter has
also recently been seen during periods of activity in \tfe\
\citep{tgd+08}.

Previously \citet{klc00} noted that the existence of AXP glitches
having properties comparable to those seen rotation-powered pulsars
was not in itself evidence for AXPs being magnetars.  This is because
the standard glitch model applies regardless of the mechanism by which
the crust slows down; i.e., glitches can occur, in principle, in an
accretion-powered pulsar. For example, a glitch was a plausible
explanation for the fractional change of $\Delta \nu/\nu \sim
4\times10^{-5}$ seen in the accreting neutron star system
KS~1947$+$300 \citep{gml04}.  However, although glitches can occur in
the case of accreting systems, since there too crust/superfluid
angular velocity lags can develop, only in the context of magnetars
has the possibility of more slowly rotating superfluid been suggested
\citep{tdw+00}.  The net spin-down in \oft\ may thus add to the
already large amount of evidence against any accretion-powered origin
for \oft.

\subsection{The Active Phase of \oft\ and Other AXP Outbursts}

In many ways, the 2006-2007 active phase of \oft\ is similar to other
phases of activity seen in AXPs: it was punctuated by short bursts,
pulse profile changes were seen, and it was accompanied by a
significant rotational anomaly.  However the \oft\ activity is unique
in one interesting way: the pulsar did not suffer a large, long-lived
pulse flux increase at the beginning of the phase.  Indeed its pulsed
flux (Fig. 9), apart from very near bursts, has remained relatively
stable.  We note that observations with focusing X-ray telescopes may
reveal some phase-averaged flux variations \citep[see][]{gdk+10};
indeed pulsed fraction has been shown to be inversely correlated with
total flux in \tfe\ \citep{tmt+05,tgd+08}, rendering pulsed flux
variations smaller relative to total variations.  Still, for \oft,
such a correlation would have to conspire to render the pulsed flux
steady. This seems unlikely, however only focusing telescope
observations can rule this out.

Other AXP radiative outbursts have been suggested as arising from
large magnetospheric twists, with associated magnetospheric currents
returning to the stellar surface and heating it, resulting in
increased X-ray emission from the source
\citep[e.g.][]{tlk02,bt07,bel09}.  Such twists are thought to
represent a release of magnetic energy and helicity from the
internally wound-up magnetic field.

On the assumption that the pulsed flux is a reasonable proxy for the
total flux for \oft, here, we find no clear evidence for significantly
increased X-ray emission on time scales longer than a few minutes.
Thus any large-scale magnetospheric twist scenario is problematic.  As
originally showed by \citet{tlk02}, the X-ray luminosity due to
returning currents reheating the surface in a significantly twisted
magnetosphere in general are comparable to that produced from internal
processes; this is clearly not observed in the 2006-7 active period of
\oft.

Instead, for this source, long-term evolution of the crust, driven
presumably internally by field decay, and resulting in multiple
unstable configurations though without any large scale magnetospheric
twists, could result in sudden cracking and local rearrangements.
This could be accompanied by bursts and profile changes, as well as
with vortex line shifting.  Why such surface motion does not produce
significant magnetospheric twists is, however, puzzling, given that
field lines are thought to be anchored in the crust; perhaps only for
large motions do field lines become sufficiently twisted for enhanced
X-ray emission to be produced.

\citet{dkg08} showed that other AXP glitches have been unaccompanied
by radiative changes.  For example, AXP \efo\ has glitched multiple
times yet its pulsed flux remains very steady.  Such radiatively
``silent'' glitches may be a result of internal activity that does not
result in significant twisting of outer magnetosphere field lines,
where as radiatively ``loud'' glitches could be those for which the
magnetosphere is impacted.  We note in fact that thus far, the data
are consistent with {\it all} AXP radiative outbursts being
accompanied by timing anomalies.  This may also be true of SGRs,
though in those cases, timing anomalies are harder to establish
because of the difficulty in achieving phase-coherent timing in
quiescence.

\section{Summary and Conclusions}
\label{sec:summary}

We have reported on anomalous X-ray pulsar \oft\ entering an active
phase which was preceded with by a long-term increase in pulsed
flux. The active phase, which commenced in 2006 March, consisted of a
timing anomaly that can be described as a net ``anti-glitch,'' that
is, a net spin-down following an initial spin-up that decayed on a
time scale of 17 days.  Following the glitch, we detected six bursts
from the pulsar, the first ever observed from this source.  Despite
10~yr of \xte\ monitoring, the bursts all occurred in the narrow time
span between 2006 April 6 and 2007 February 7.
The sixth and largest burst had a unusual spectrum that cannot be
described by any simple continuum model.  Rather, it can be well fit
if a broad spectral feature at $\sim$14~keV is included, as has
been reported in other AXP bursts, as well as a narrow feature at
$\sim$8~keV.  The pulse profile of the source changed
from double- to triple- peaked near the bursts, and underwent
considerable evolution otherwise.  At the burst epochs the relative
intensity of the three peaks significantly varied.  The pulse profile
is now relaxing to its pre-active phase morphology.  Most aspects of
\oft's emission changed during the active phase, with the notable
exception of the pulsed flux (except near bursts).  This argues
against this event being associated with a sudden magnetospheric
twist, as has been invoked for other AXP activity, and is suggestive
of crustal evolution driven internally by the large magnetic field,
though without significant magnetospheric twisting.  We suggest that
other, radiatively silent AXP glitches have a similar origin, whereas
radiatively loud AXP timing events occur when the crustal motions
cannot avoid significant twisting of the magnetic field lines.

\acknowledgements We thank P.~M.~Woods and A.~M.~Beloborodov for
useful discussions.  This research has made use of data obtained
through the High Energy Astrophysics Science Archive Research Center
Online Service, provided by the NASA/Goddard Space Flight Center. This
work has been supported by an NSERC Discovery Grant, the Canadian
Institute for Advanced Research, and Le Fonds Qu\'eb\'ecois de la
Recherche sur la Nature et les Technologies, by the Canada Research
Chairs program, and by the Lorne Trottier Chair in Observational
Astrophysics.


\end{document}